\documentclass[10pt]{article}

\usepackage{amsmath}
\usepackage{amsfonts}
\usepackage{amssymb}
\usepackage{graphicx}
\usepackage{setspace}
\usepackage{authblk}
\usepackage{cite}
\usepackage{setspace}

\usepackage[linktocpage=true]{hyperref}
\usepackage{float}
\usepackage[dvipsnames]{xcolor}
\usepackage{hyperref}
\hypersetup{colorlinks,linkcolor={BrickRed},citecolor={RoyalBlue},urlcolor={RoyalBlue},filecolor={Turquoise}}

\begin{document}
	
	\title{\textbf{Dark Matter perspective of Left-Right symmetric gauge model} }
	\author{Sanchari Bhattacharyya\footnote{sanchari1192@gmail.com}\;}
	\author{\;Anindya Datta\footnote{adphys@caluniv.ac.in}}
	\affil{\emph{Department of Physics, University of Calcutta} \\ \emph{92, Acharya Prafulla Chandra Road, Kolkata 700009}}
	\date{}
	\maketitle
	
	\vskip 2cm
	
		\begin{abstract}
		We consider an incarnation of left-right symmetric model with a local gauge symmetry of $SU(3)_C \otimes SU(2)_L \otimes U(1)_L \otimes SU(2)_R \otimes U(1)_R$. Heavy scalars and fermions present in the {\bf 27} of $E_6$ are included in the matter sector along with the Standard Model (SM) fermions. Two such color singlet fermions, $N$ and $L_S$, transforming as bi-doublet and singlet under $SU(2)$s respectively, can be potential candidates for Dark Matter (DM). Assignment of $U(1)$ charges for the matter fields restricts some of the exotic fermions to interact with the SM fermions. We study in some details the prospect of such fermionic dark matters by calculating relic densities and direct detection cross-sections by treating both these particles as relics. In such a two component Dark Matter scenario, $L_S$ having smaller interaction with the SM, will dominantly contribute to relic density. However, it cannot be detected at earth bound experiments with their present sensitivity. On the contrary, $N$ having higher rate of interaction with the SM particle has too large annihilation cross-section thus contributes very little to relic density. In fact, its interaction with the SM is too high such that $N$-nucleon cross-section for a wide range of $N$ mass is higher than the experimental limits from XENON, LUX or PICO. However, such a high $N$-nucleon cross-section can be tamed by assuming additional dimension-6 operators involving $N$ and SM quarks. We derive limits on the strengths of such interactions from experimental data.

	\end{abstract}
	
	
	\hrulefill

	\section{Introduction}
	In recent times, several cosmological parameters have been measured experimentally at unprecedented precision. One such example is the experimental data from several independent experiments, mounting to unavoidable evidence in support of a non-luminous matter, more commonly known as Dark Matter (DM) present in the entire Universe. The measurement from the PLANCK \cite{Planck} reveals that luminous matter constitutes only 4-6\% of the energy density of the Universe whereas almost 26\% of it is accounted by DM, whose exact nature is still an enigma to us. This partitioning of energy density of the Universe has also been in agreement with the measurement of CMBR anisotropy from WMAP \cite{WMAP}. Indirect evidences in support of existence of DM have also been gathered more recently, from satellite based experiments like AMS \cite{AMS}, PAMELA \cite{Pamela} and Fermi-LAT \cite{Fermi1,Fermi2}.
	
	On the particle physics front, discovery of the Higgs boson \cite{higgs-discovery} and the ongoing measurement of its properties at the Large Hadron Collider (LHC) experiment once again has firmly established the validity of the Standard Model (SM) at TeV scale. However, the absence of a viable DM candidate and massive neutrinos are major shortfalls of the so far successful model of interactions among elementary particles and fundamental forces. Although, in earlier times, weakly interacting neutrinos have been thought to be the candidates for the DM, with advent of more and more precise cosmological data, neutrinos are disfavoured.

	The demand for a viable DM candidate has been one of the main motivations to look beyond the SM (BSM). In the post Higgs-boson discovery era, pursuit for a dark matter is the prime aim for experimental and theoretical front. Supersymmetric (SUSY) \cite{SUSY-DM,susy} and extra-dimensional models \cite{extrad} have been so far very popular and thus they have been extensively investigated BSM scenarios. Imposition of a $Z_2$-symmetry ($R$-parity for SUSY and $KK$-parity for extra-dimensional models) on the action, ensures the stability of the lightest SUSY (KK) particle which can be a viable DM candidate. However, non-observation of any tele-tale signature of any kind of new physics from the LHC, only has pushed the lower mass limits of SUSY or KK -particles in the TeV range\cite{LHC limits}. Several other variants of DM models have also been proposed. Little Higgs model \cite{little-higgs}, left-right symmetric models \cite{LRDM}, models with extended scalar sector \cite{extended scalar} or an $U(1)$ extended SM \cite{extrau1} are notable among them.
	
	There is a class of left-right symmetric models, where scalar as well as fermionic dark matter have been studied \cite{LRDM}. In most cases the minimal LR symmetric models are extended by adding scalar or fermionic multiplets who can be potential candidate for a dark matter. In some cases, the right handed neutrino in LR models has been investigated as a suitable DM candiadte. 
	
	In the present article, in pursuit of a DM, we will turn our attention to a model with local gauge invariance under $SU(3)_C \otimes SU(2)_L \otimes U(1)_L \otimes SU(2)_R \otimes U(1)_R$ ($32121$). This particular gauge group would result from a two step breakdown of $E_6$ \cite{E6}. However, in this work we will not be interested in any of the effects of $E_6$ that might be carried down to EW scale via a renormalisation effect. To have a viable DM candidate, we take the resort from the particle content of {\bf 27} dimensional representation of $E_6$ without restricting ourselves to the rules of $E_6$ breaking in determining the hyperchages of the new particles those have been augmented in our model. Assignments of hyper-charges (for these fermions, from the consideration of anomaly cancellation) restrict SM fermions to have couplings with the exotic fermions and some of the Higgs bosons. This in turn pave the way for some of the fermions (with zero electric charge) to be viable DM candidates.
	A similar variant of the model of our interest model and its phenomenology in the context of the LHC has been discussed in a previous article \cite{32121}.	
	In this article, we will concentrate on the viable candidates of DM in 32121 model, whether the masses and interactions of such particles are in the right ballpark to satisfy the relic density and direct detection limits obtained from experiments.
	
	The idea of fermionic dark matter has been exclusively investigated. Previously many authors \cite{extrau1, fermion-DM, axion, boltzman} have studied the possibility of a DM which has spin- $1\over2$, both Dirac and Majorana in nature with their masses varying from sub-MeV scale to TeV scale. However in all of these studies, the SM has been augmented by such fermionic fields, which couple to the SM via either a scalar or a gauge boson not present in the SM. Couplings of such exotic particles to the SM have been restricted by imposing some extra discrete symmetries on the action or in some cases such an extra symmetry is a remnant of breakdown of some bigger symmetry already existing in the action \cite{axion}. 
	In most of these studies, the spin- $1\over2$ DM, is accompanied by another relic particle also having restricted coupling to SM particles.
	However, novelty of our analysis lies in the presence of a pair of spin- $1\over 2$ relic particles, one Dirac and the other Majorana in nature along with few other fermions, arising from the full $\textbf{27}$-plet of $E_6$.
	We have already mentioned that we need not to impose any discrete symmetry to restrict the couplings of the DM candidates to the SM fields. Such couplings have been automatically not allowed from the $U(1)_{L,R}$ assignments of such fermions.
	It is important to mention that few of these exotic fermions (which have electromagnetic/weak charges) also play a crucial role in co-annihilation processes thus indirectly contribute to relic density. This is certainly a hallmark of a more {\it complete} BSM like MSSM, where the extra particle fields arise to fulfill the conditions of full symmetry of the theory. Furthermore, in our case, gauge symmetry allows a Higgs mediated interaction between two DM candidates, which we will see, plays a crucial role in determining the relic density. 	
	
	In the next section (Section \ref{sec2}), we will briefly review the $32121$ model with emphasis on the possible DM candidates and their interactions. Section \ref{sec3} will deal with the issue of direct detection of the DM on earth bound experiments. We will estimate the DM nucleon cross-section and compare the results with experimental data from XENON, LUX and PICO. This will be followed 
	by a very brief discussion of the standard route to relic abundance calculation starting from the interactions and identification of annihilation and co-annihilation channels. We will also present the main results of our analysis in this section. Finally, we conclude in section \ref{sec5}.

	\section{Description of 32121 Model}
	\label{sec2}
	
	We are interested in a left-right (LR) symmetric gauge group $SU(3)_C \otimes SU(2)_L \otimes U(1)_L \otimes SU(2)_R \otimes U(1)_R$. This can be a result of two step breaking of $E_6$. We are not interested in the exact mechanism of this breaking chain, at this moment. Instead our focus will be on some of these electrically neutral leptons and their interactions with other particles. This choice of fermions, with charge assignments listed in Table \ref{table1}. 
	
	Gauge bosons present in this model automatically follow from the gauge group. The matter and gauge fields which are present in our model including the Higgs multiplets along with the gauge quantum numbers instrumental in breaking down $SU(3)_C \otimes SU(2)_L \otimes U(1)_L \otimes SU(2)_R \otimes U(1)_R$ to the SM gauge group are listed in Table \ref{table1}. Electric charge, $Q$ is defined through the relation, $Q = T_{3L} + T_{3R} + Y_L/2 + Y_R/2$. $L$ and $R$ stand for left and right repectively.
	
	\begin{table}[h!]
		\centering
		\begin{tabular}{|c|c|c|c|c|c|c|}
			\hline \hline
			& & $3_C$ & $2_L$ & $2_R$ & $1_L$ & $1_R$ \\ [1.0ex]
			\hline 
			& $L_L$ & $1$ & $2$ & $1$ & $-1/6$ & $-1/3$ \\ [1.0ex]
			& $\bar{L}_R$ & $1$ & $1$ & $2$ & $1/3$ & $1/6$ \\ [1.0ex]
			& $\bar{L}_B$ & $1$ & $2$ & $2$ & $-1/6$ & $1/6$ \\ [1.0ex]
			Fermions & $\bar{l}_S$ & $1$ & $1$ & $1$ & $1/3$ & $-1/3$ \\ [1.0ex]
			& $Q_L$ & $3$ & $2$ & $1$ & $1/6$ & $0$ \\ [1.0ex]
			& $\bar{Q}_R$ & $\bar{3}$ & $1$ & $2$ & $0$ & $-1/6$ \\ [1.0ex]
			& $\bar{Q}_{LS}$ & $\bar{3}$ & $1$ & $1$ & $-1/3$ & $0$ \\ [1.0ex]
			& $Q_{RS}$ & $3$ & $1$ & $1$ & $0$ & $1/3$ \\ [1.5ex]
			\hline				
			& $\Phi_B$ & $1$ & $2$ & $2$ & $1/6$ & $-1/6$ \\ [1.0ex]
			Higgs & $\Phi_L$ & $1$ & $2$ & $1$ & $1/4$ & $1/4$ \\ [1.0ex]
			Bosons& $\Phi_R$ & $1$ & $1$ & $2$ & $-1/4$ & $-1/4$ \\ [1.0ex]
			& $\Phi_S$ & $1$ & $1$ & $1$ & $-1/3$ & $1/3$ \\ [1.5ex]
			\hline				
			& $G^i ,\; i=1,...,8$& $8$ & $1$ & $1$ & $0$ & $0$ \\ [1.0ex]
			& $W^i_{L}, i=1,2,3$ & $1$ & $3$ & $1$ & $0$ & $0$ \\ [1.0ex]
			Gauge bosons & $W^i_{R}, i=1,2,3$ & $1$ & $1$ & $3$ & $0$ & $0$ \\ [1.0ex]
			& $B_L$ & $1$ & $1$ & $1$ & $0$ & $0$ \\ [1.0ex]
			& $B_R$ & $1$ & $1$ & $1$ & $0$ & $0$ \\ 
			\hline
		\end{tabular}
		\caption{Fermions and Bosons in $32121$ model with their respective gauge quantum numbers}
		\label{table1}
	\end{table}

	\subsection{Brief Description of the Scalar, Gauge and Fermion sector}
	\label{subsec21}
	
	{\bf Scalar sector:} The scalar sector of the $32121$ model contains one Higgs bi-doublet ($\Phi_B$), one left-handed ($\Phi_L$), one right-handed ($\Phi_R$) weak doublets and a singlet Higgs boson ($\Phi_S$) with non-zero $U(1)_{L,R}$ charges. These scalars arise from the (\textbf{1}, \textbf{3}, $\bar{\textbf{3}}$) representation of $\left[ SU(3)\right]^3$. For a complete symmetry breaking mechanism from $32121 \longrightarrow SU(3)_C \otimes SU(2)_L \otimes U(1)_Y \longrightarrow SU(3)_C \otimes U(1)_{EM}$, one needs a bi-doublet scalar field $\Phi_B$, two doublets $\Phi_L$ and $\Phi_R$ along with a $SU(2)$ singlet Higgs field $\Phi_S$. Five neutral CP-even scalars ($h^0$, $h_2^0$, $h_L^0$, $H_R^0$, $H_S^0$), two CP-odd scalars ($\xi_2^0$, $\xi_L^0$) and two charged scalars ($H_1^\pm$, $H_L^\pm$) are left after EWSB. One from the CP-even sector ($h^0$) is identified with the SM Higgs boson. Two neutral scalars, $h_L^0$ and $\xi_L^0$ (originating from $\Phi_L$) having nominal interaction to other SM particles, could possibly be relics and we will see whether they could satisfy the limits of relic density and scattering cross-section from direct detection of DM. \\
	
	{\bf Gauge sector:} The electro-weak gauge sector of $32121$ model consists of two charged gauge bosons $W$, $W'$ and four neutral gauge bosons. Two of them can be identified with the SM-like $Z$ boson and photon. The remaining two massive neutral gauge fields are identified as $Z'$ and $A'$ where the appearance of $A'$ is the consequence of the extra local $U(1)$ symmetry. Their masses and interactions are governed by 4 gauge coupling constants $g_{2L}$, $g_{2R}$, $g_{1L}$ and $g_{1R}$ and non-zero vacuum expectation values (vevs) of the scalar fields.
	Following the symmetry breaking pattern from $SU(2)_R \otimes U(1)_L \otimes U(1)_R$ to $U(1)_Y$, one can have,
	\begin{equation}
	\label{GCValue1}
	\dfrac{1}{g_Y^2} = \dfrac{1}{g_{2R}^2} + \dfrac{1}{g_{1L}^2} + \dfrac{1}{g_{1R}^2}
	\end{equation} where $g_Y$ is the gauge coupling constant corresponding to $U(1)_Y$. $g_{2L}$ is assumed to be equal to $SU(2)_L$ gauge coupling constant, $g$ of SM.
	However, in order to keep our Lagrangian manifestly LR symmetric, we assume $g_{2L}=g_{2R} = g$ and $g_{1L}=g_{1R}$. All our analysis presented in the following will be based on this assumption. The vevs of the scalar fields can be constrained from below from the experimental lower limit of heavy gauge boson masses. For a detailed discussion about the gauge sector we refer the reader to an earlier work \cite{32121}.\\
	
	{\bf Fermion sector:} 
	The chiral components of the fermions of 32121 model are listed below. \begin{eqnarray}
		L_{L} &=& \begin{pmatrix}
		\nu_{L} \\ e_{L}
		\end{pmatrix}, \hspace{1.2cm} \hspace{1.3cm} L_{R} = \begin{pmatrix}
		\nu_{R} \\ e_{R}
		\end{pmatrix} \nonumber \\
		Q_{L} &=& \begin{pmatrix}
		u_{L} \\ d_{L}
		\end{pmatrix}, \hspace{2.4cm} Q_{R} = \begin{pmatrix}
		u_{R} \\ d_{R}
		\end{pmatrix} \nonumber \\
		Q_{LS} &=& q_{SL}, \hspace{0.5cm} Q_{RS} = q_{SR}, \hspace{0.5cm} l_S \hspace{0.5cm} \mbox{and,}\nonumber \\
		L_B &=& \begin{pmatrix}
		N_1 & E_1 \\ E_2 & N_2
		\end{pmatrix}  \hspace{0.3cm} \mbox{and} \hspace{0.7cm}
		\tilde{L}_B = \begin{pmatrix}
		N_2^c & E_2^c \\ E_1^c & N_1^c
		\end{pmatrix}
	\end{eqnarray}
		
	Here, $L_L,L_R,Q_L,Q_R$ comprises of 16 fermions among which 15 are present in SM. $N_1$ and $N_2$ are neutral heavy leptons while $E_1$ and $E_2$ are singly charged heavy leptons.
	
	Apart from SM fermions, the model under consideration contains a right-handed neutrino $\nu_R$, an $SU(2)_L \times SU(2)_R$ singlet quark $q_S$, (a Dirac fermion constructed out of Weyl fermions $Q_{LS}$ and $Q_{RS}$) and $E$ ($N$) transforming non-trivially under $SU(2)_L \times SU(2)_R$ (a Dirac fermion constructed out of Weyl fermions $E_1$ ($N_1$) and $E_2^c$ ($N_2^c$)). The gauge quantum numbers of the fermions used in our analysis have been listed in Table \ref{table1}. Apart from these, we also have a $SU(2)$ singlet charge neutral lepton (Weyl spinor) $l_S$ in the spectra. $L_S$ is a Majorana fermion constructed out of $l_S$ and $l_S^c$. Both $N$ and $L_S$ can be viable dark matter particles.
	
	Fermions get their masses via the interactions with Higgs fields. The relevant Yukawa Lagrangian is noted below.
	\begin{equation}
	\label{LRYukawa}
	\mathcal{L}_{Yukawa} = \mathcal{L}_{Y4} + \mathcal{L}_{Y5}
	\end{equation} where,
	\begin{eqnarray}
	\label{LRYukawa4}
	\mathcal{L}_{Y4} &=& y_{qij}~ \bar{Q}_{iL} \Phi_B Q_{jR} + \tilde{y}_{qij}~ \bar{Q}_{iR} \tilde{\Phi}_B Q_{jL} + y_{lij}~ \bar{L}_{iL} \Phi_B L_{jR} + \tilde{y}_{lij}~ \bar{L}_{iR} \tilde{\Phi}_B L_{jL} \nonumber \\ 
	&+& y_{sij}~ \bar{Q}_{iLS} \Phi_S Q_{jRS} + y_{LBij} \; Tr \left[ \bar{L}_{iB} \tilde{L}_{jB} \right] \Phi_S^* + y_{BBij}\; Tr \left[ \bar L_{iB} \tilde \Phi_B \right] l_{jS}^c + H.c .
	\end{eqnarray} 
	\begin{eqnarray}
	\label{LRYukawa5}
	\mathcal{L}_{Y5} &=& \frac{1}{\Lambda} \left[~ y_{LSij}~ \bar{l}_{iS} l_{jS}^c \Phi_S \Phi_S + y_{SSij}~ Tr \left[ \bar{L}_{iB} \Phi_B \right] l_{jS} \Phi_S^* + y_{qSBij}~ \bar{Q}_{iLS} Tr \left[\Phi_B^\dagger \tilde{\Phi}_B \right] Q_{jRS} \right]+ H.c	
	\end{eqnarray}
	where, $i,j = 1,2,3$ are generation numbers and $y$(s) are Yukawa coupling constants. $\Phi_S^*$ is complex conjugate of $\Phi_S$, $\tilde{\Phi}_B = \sigma_2 \Phi_B^* \sigma_2$. $\mathcal{L}_{Y4}$ and $\mathcal{L}_{Y5}$ comprise of the dimension-4 and 5 operators respectively depicting interactions between the fermions and scalars.
	
	Yukawa interactions which generate masses for the SM fermions are noted in the first line of Eq. \ref{LRYukawa4}. 
	It is interesting to note, one can only write Yukawa interactions of the SM fermions to the bi-doublet Higgs $\Phi_B$ while exotic fermions can get their masses via their interactions with the singlet Higgs field $\Phi_S$ (see the second line of Eq. \ref{LRYukawa4}) only. This is a consequence of assignments of $U(1)_{L,R}$ charges of the fermions and scalars. A notable advantage of absence of such interactions between exotic and SM fermions facilitate us to choose $N$ and/or $L_S$ to be a DM candidate. One does not have to impose any discrete symmetry on the fields to prevent the DM particle from decaying to a pair of lighter SM particles \cite{Triparno-Rinku}.
	
	The Yukawa coupling matrices are considered as non-diagonal. 
	It is important to mention that there are no such term present in the Yukawa Lagrangian that leads to mixing among exotic and SM fermions. So the mass matrices of SM and exotic fermions can be brought into diagonal form independent of each other. This has another important consequence namely the identification of the light neutrino species. The heavy neutral leptons ($N$ and $L_S$) are decoupled from the neutral leptons arising from $L_L$ and $L_R$. So in principle, by appropriately choosing $y_{ij}$ one can write Dirac masses for light neutrino specie. However, without choosing small Yukawa couplings for light SM like neutrinos one turns to other aesthetically acceptable mechanisms \cite{see-saw} which have been widely discussed in literature. But we reiterate that the presence of exotic heavy neutral leptons (the DM candidates in our case) in no way affect the identification of light neutrino species in this model. Thus while considering the phenomenology of $N$ and $L_S$, we may use their physical masses as the free parameters of the analysis.
	
	It is important to note, a dimension-4 mass term for the singlet lepton $l_S$ (a Weyl spinor) cannot be written as it transforms non-trivially under $U(1)_{L,R}$. But we are able write a dimension-5 operator, which in turn generates mass for $l_S$. To generate a mass using Higgs mechanism, one must employ a fermion from a multiplet of $E_6$ other than {\bf 27} (e.g., \textbf{78} representation). So $\Lambda$ may be identified with the mass of such a fermion. The last term in Eq. \ref{LRYukawa5}, signifies a dimension-5 mass term for the singlet exotic quarks. However, a dimension-4 mass term for the same, has already been written in Eq. \ref{LRYukawa4} and we will not consider any effect of this dimension-5 term in our analysis. Similarly, a term proportional to $y_{SSij}$ will not be considered any more, as it has the same consequences as the dimension-4 term proportional to $y_{BBij}$.	
	
	The last term in Eq. \ref{LRYukawa4} seeks our attention. It introduces a mixing between the singlet lepton ($l_S$) and the charge neutral lepton ($N$) via the bi-doublet Higgs boson $\Phi_B$. Neglecting the effects of other dimension-5 operators except the mass of $l_S$, mass terms for the heavy neutral fermions in the Yukawa Lagrangian can be written as follows.
	\begin{eqnarray}
	\begin{pmatrix}
	\bar{N_1} & \bar{l}_S
	\end{pmatrix} 
	\begin{pmatrix}
	\sqrt{2} y_{LB} v_S & 0 \\
	\dfrac{y_{BB}}{\sqrt{2}} k_1 & \dfrac{y_{LS}}{2 \Lambda} v_S^2 
	\end{pmatrix}
	\begin{pmatrix}
	N_2^c \\ l_S^c
	\end{pmatrix} + H.c.
	\label{Nmass}
	\end{eqnarray}
	$k_1$ and $v_S$ are the vevs of $\Phi_B$ and $\Phi_S$ respectively. Being proportional to $k_2$, the 12 element of the above matrix vanishes when $k_2=0$. As thoroughly discussed in \cite{32121}, value of $k_1$ has been fixed to 246 GeV from SM $W$ boson mass. A lower limit on $v_S \sim 12.6$ TeV has been derived from the lower mass limit of $A' (\sim 3.5 ~\mbox{TeV})$.
	Given the fact that $k_1 << v_S$, the terms proportional to $y_{BB}$ will introduce a nominal mixing between $N$ and $L_S$.
	
	This mixing will introduce a mass difference between the physical eigenstates corresponding to $N$ and $L_S$. A tiny value of $y_{BB}$ can lower this mass difference below the KeV level making any of the  above physical states, a stable DM candidate. However, we shall see in next section that, we need a sizeable $y_{BB}$ (of the order of 0.2 or higher) which  makes the above mass difference of the order of  few tens of GeV. In Table \ref{table2}, we  note down the mass difference between the eigenstates  corresponding to $N$ and $L_S$ following the diagonalisation of Eq. \ref{Nmass}, for different values of $y_{BB}$ allowed by Planck limit. With these choices of the parameters the mass of $N$ will be always be lower than that of $L_S$. However, we must ensure that for a given mass difference between $L_S$ and $N$, the decay lifetime of the heavier must be much greater than the lifetime of the Universe. For the benchmark points mentioned in Table \ref{table2}, (with mass differences of the order of $40$ GeV) the decay lifetime of the heavier DM candidate will be $\sim 10^{-10}$ s (with $\Gamma \sim 10^{-15}$ GeV) which is much much smaller than the required lifetime ($> 10^{22}$ s \cite{W-mixing}). A simple calculation would reveal that such a large lifetime could only be achieved only when the aforementioned mass difference is less than few hundreds of KeVs. So in order to deal with such a situation, one may further notice the fact that, $N$ and $L_S$ has different gauge transformation properties which results into different radiative corrections to their masses. In general, at the leading order, radiative corrections to $N$ or $L_S$ masses will be grossly proportional to $\left( \frac{g_\ast}{4 \pi} \right)^2 m_{N (L_S)} \; \log \left ( \frac {m_X^2 } { m ^2 _{N (L_S) }} \right)$ where $g_*$ is the coupling between $N (L_S)$ and the particle $X$ (possibly a gauge boson or a scalar) forming the loop. 
	Assuming, $X$ to be one of the heavy particles like $Z'$, $W'$ or $A'$, or heavy Higgs bosons in the model (masses at the order of 5 TeV), the above expression amounts to $4 - 9$ GeV. One must remember that $N$ always will receive extra contributions from $SU(2)$ gauge bosons apart from $U(1)$ gauge bosons. So adding, contributions from all such diagrams possibly compensate the mass gap generated by the presence of $y_{BB}$ term in the Yukawa Lagrangian and bring the mass gap below a few hundreds of KeV. However, the full estimation of the radiative correction of the masses is beyond the scope of this article.		
		\begin{table}[H]
		\centering
		\begin{tabular}{|c|c|c|c|}
			\hline \hline	
			$\sqrt{2} y_{LB} v_S$  & $\dfrac{y_{LS}}{2 \Lambda} v_S^2$  & Allowed $y_{BB}$ & $\vert m_N - m_{L_S} \vert$  \\ [1.0ex]
			[TeV] & [TeV] & &  [TeV]\\ [1.0ex]
			\hline \hline				
			0.99 & 1 & 0.220   & 0.03955  \\ [1.0ex]
			1.99 & 2 & 0.229  & 0.04107 \\ [1.0ex]
			2.99 & 3 & 0.220   & 0.03956 \\ [1.0ex]
			3.99 & 4 & 0.223  & 0.04006 \\ [1.0ex]
			\hline \hline
		\end{tabular}
		\caption{The mass differences ($\vert m_N - m_{L_S} \vert$) at tree level between the physical eigenstates corresponding to $N$ and $L_S$ for allowed (from Planck limit) value of $y_{BB}$.}
		\label{table2}
	\end{table}
	
	For all practical purpose, $N$ and $L_S$ are the physical eigenstates\footnote{To ascertain this, we have kept the diagonal 
	elements of the mass matrix in Eq. \ref{Nmass} slightly different from each other.}. The corresponding eigenvalues will be used as free parameters in our analysis without any loss of generality. However, we will see that this particular interaction will play a crucial role while evaluating the relic density in case of a two-component DM scenario comprising of $N$ and $L_S$.
	
	$N$ and $E$ get their masses from the same source\footnote{A lower mass limit in the ballpark of a TeV ($>$ 1.089 TeV), on $m_{E^\pm}$ has been derived in \cite{32121} from the exisiting experimental result on long lived charged particle search at the LHC.}. As long as $y_{BB} =0$, they are mass degenerate and have masses equal to $\sqrt{2} y_{LB} v_S$. For a non-zero, $y_{BB}$ (see Eq. \ref{Nmass}), $m_N$ becomes lighter than $E^\pm$ for a wide range of $y_{BB}$. 
	Furthermore, $E^\pm$ being electrically charged, receives contribution to its mass from radiative corrections. These corrections are beyond the scope of this study.
	However, it is necessary for $E$ to be heavier than $N$ so that it can decay and thus does not contribute to the relic. The necessary mass difference between $N$ and $E$ could be thought to be generated via aforementioned mechanisms.

	\subsection{Dark Matter Candidtaes in 32121}
	
	With a careful study of the particle sector of 32121 model, one can identify a number of new particles which could be eligible candidates for the dark matter. A suitable DM candidate must,
	
	\begin{itemize}
		\item be stable (or for decaying DM, lifetime $>$ age of the Universe) and possibly charge neutral.
		\item satisfy the limits on relic abundance of the Universe in present epoch.
		\item agree with the upper limits on DM-nucleon scattering cross-section obtained from the direct detection experiments.
		\item agree with the limits on decay branching ratio \cite{Higgs-invisible} of SM-Higgs to invisibles if the relic couples to the SM Higgs boson.
	\end{itemize}
	
	In the following we discuss quantitatively the prospects of some of the suitable DM candidates in this model.
	\vspace{2mm}
	
	\textbf{DM candidates arising from scalar sector:}
	$h^0_L$ and $\xi^0_L$ are degenerate in mass. Due to their mass degeneracy direct detection cross-section of such scalar dark matter is well above the limit coming from an experiment like XENON \cite{scalar-DM1}. By the same token of argument their annihilation cross-section to the SM particles is very high resulting into a very tiny relic density well below the PLANCK limit. We abandon the case of scalar dark matters and focus into fermionic dark matter in our following analysis.
	
	\vspace{2mm}
	
	\textbf{DM candidates arising from fermion sector:}
	 Eq. \ref{LRYukawa} reveals that the mixing of exotic fermions with SM fermions are prohibited and exotic fermions cannot decay to a pair of SM particles. In such a scenario, the two neutral exotic leptons $N$ and $L_S$ can also be eligible DM candidates.
	
	Now, in the other scenario when the last term in Eq. \ref{LRYukawa} is switched on, some new interactions come into play which may make any one of $L_S$ or $N$ to be unstable. For example, turning on $y_{BB}$ gives rise to some interactions among $N$, $l_S, E^\pm$ and the neutral ($h^0$, $\xi_2 ^0$, $h_2^0$, $H_S^0$) and charged scalar ($H_1^\pm$). In such a case, the mass difference between $N$ and $L_S$ plays an important role in co-annihilation of the DM.
	
	 In the next section, we will first calculate the direct detection cross-section of $N$ and $L_S$ seperately followed by an estimation of relic abundance.

	\section{Study on the Dark Matter in 32121 Model}
	\label{sec3}
	
	In the previous section, we have identified the particles which can satisfy the DM criteria. In this section, we shall concentrate on detail analysis of the aforementioned WIMP-like DM candidates. We will estimate the DM-nucleon scattering cross-sections and compare the results with the experimental data by XENON collaboration \cite{direct-detection} and other experiments like LUX \cite{lux} and PICO \cite{pico}. Next, we will analyse whether they can satisfy present DM relic abundance of the Universe \cite{Planck}. 		
	
	\subsection{Direct Detection of Dark Matter}
	\label{subsec31}
	
	To start with, we concentrate on the elastic scattering between the nucleus and the relic particle. Experiments like XENON \cite{direct-detection}, LUX \cite{lux}, PANDA \cite{panda} provide upper limits on the DM-nucleon scattering cross-section. We have considered two fermionic dark matter candidates, namely $N$, a Dirac fermion and $L_S$ which is a Majorana-like fermion. 
	
	\vspace{5mm}
	
\begin{figure}[h!]
		\begin{center}
			\includegraphics[width=8cm, height=4cm]{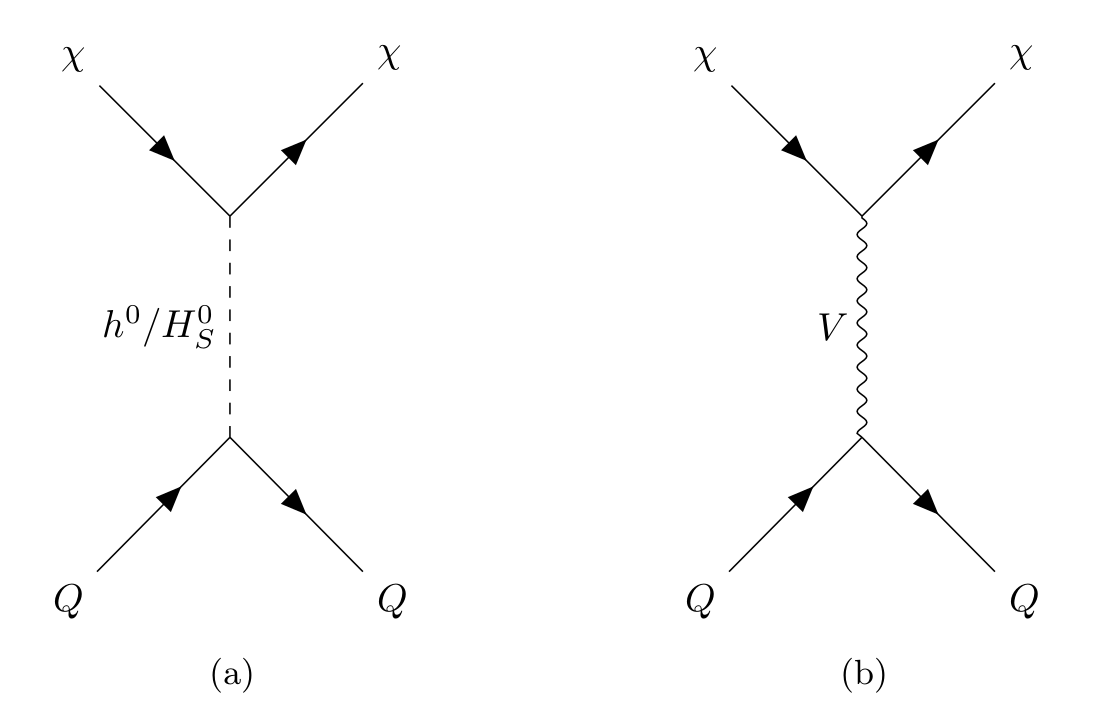}
		\end{center}
		\caption{The tree-level Feynman diagrams corresponding to the operators in Eq. \ref{NDDs} (a) and Eq. \ref{NDDv} (b). $V$ represents a massive neutral gauge boson of $32121$ model.}
		\label{NDDdiag}
	\end{figure}

	For a generic fermionic DM, $\chi$ (which can be identified with $N$ or $L_S$ in our model), the quark-level effective operators, responsible for its direct detection are the following,
	\begin{eqnarray}
	\label{NDDs}
	\mathcal{C}_s \left(\bar{\chi} (g_{\chi s}+ig_{\chi p}\gamma^5) \chi\right) \left(\bar{Q} (g_s+ig_p\gamma^5) Q\right) \\
	\label{NDDv}
	\mathcal{C}_v \left(\bar{\chi} \gamma^{\mu}(g_{\chi V}+g_{\chi A} \gamma^5) \chi\right) \left(\bar{Q} \gamma_{\mu}(g_{V}+g_{A}\gamma^5) Q\right) 
	\end{eqnarray}
	Here, $Q$ is a SM-quark field. Eq. \ref{NDDs} represents the case where a scalar particle acts as a mediator. $g_{\chi s(p)}$ is the coupling between DM and mediating scalar (pseudo-scalar) while $g_{s(p)}$ are couplings between the mediator scalar (pseudo-scalar) with a pair of SM-quarks. 
	Similar definition holds for the couplings in Eq. \ref{NDDv} except the fact that here the mediator is one of the heavy neutral gauge bosons of 32121 model. $V (A)$ stands for vector (axial vector) mediation in such a case. $\mathcal{C} ^{-1}_{s,v}$ is connected to the square of the mass of the scalar or vector mediator (see Fig. \ref{NDDdiag}).
	
	Fig. \ref{NDDdiag} shows the tree-level Feynman diagrams which will lead to the aforementioned effective operators in Eqs. \ref{NDDs}, \ref{NDDv}. Following Eq. \ref{LRYukawa}, considering the parameter space mentioned in \cite{32121} (where mixing between $H_R^0$ and $h^0/H_S^0$ is almost vanishing) it is clear that the mediating scalar could be either $h^0$ or $H_S^0$. 
	
	The Higgs mediated diagram in Fig. \ref{NDDdiag} (a) will be suppressed with respect to the gauge mediated diagram due to the following reasons. Primarily, Higgs couplings to the SM quarks are proportional to the quark masses thus are small. Secondly, the dark matter ($N$ or $L_S$) scatters off the quarks via these interactions to $H_S^0$ whose admixture in the relevant Higgs eigenstate is very small \cite{32121}. Spin-dependent DM-nucleon cross section which follows from Eq. \ref{NDDs} will be further suppressed by the powers of DM speed. Consequently contributions from such diagrams to total cross-sections will be negligible.
	
	Eq. \ref{NDDv} points towards spin-1 vector boson mediated DM-quark scattering corresponding to the diagram (b) in Fig. \ref{NDDdiag}. It results into both spin-independent and spin-dependent cross-sections depending on the vector and axial vector-like couplings the associated particles have. The vector mediated spin-independent cross-section will be,
	\begin{equation}
	\left(\sigma_{SI}\right)_V \simeq \dfrac{g_{\chi V}^2 \mu_{\chi N}^2}{\pi m_V^4} \left[\tilde{f}_p Z + \tilde{f}_n (A-Z)\right]^2
	\end{equation}
	$\tilde{f}_p$ and $\tilde{f}_n$ are dimensionless quantities, involving the couplings of the quarks with the mediating gauge bosons, $\mu_{\chi N}$ is the reduced mass of the WIMP-nucleus system \cite{DDref}.
	
	The spin-dependent cross-section can also be expressed as the following,	
	\begin{equation}
	\label{Nsigma_SD}
	\left(\sigma_{SD}\right)_V \simeq \dfrac{4 g_{\chi A}^2 \mu_{\chi N}^2}{\pi m_V^4} J_N (J_N+1) \left[\dfrac{\langle S_p \rangle}{J_N} \tilde{a}_p + \dfrac{\langle S_n \rangle}{J_N}\tilde{a}_n \right]^2
	\end{equation}
	$\tilde{a}_{p,n}$ are dimensionless quantities, $J_N$ is the spin of the nucleus, $\langle S_{p,n} \rangle$ are the average spin of the nucleons \cite{DDref}.
	
	Before we quote any numerical values of the cross-sections, let us quote the numerical values of the input parameters that have been used. The only model parameters that creep in the SD and SI cross-sections are the masses of the DM and the masses of the gauge bosons that facilitate the DM-nucleon interaction. We have already pointed out that Higgs mediated DM-nucleon interactions can be neglected. DM-nucleon interaction can be mediated by heavy neutral gauge bosons $Z'$ and $A'$ apart from the SM-$Z$-boson. $L_S$ does not couple to SM Z boson. In case of spin-independent scattering cross-section, coupling of $N$ to SM Z boson vanishes. But for the spin-dependent case, $N$ has a non-zero coupling to SM Z boson. While evaluating the aforementioned cross-sections, we have set $A'$ and $Z'$ masses at their lower limits of $3.5$ TeV and $5.9$ TeV respectively, derived from experimental data \cite{32121}. $g_{\chi V} = g_{1L} Y_L c_{3j} + g_{1R} Y_R c_{4j}$ and $g_{\chi A} = g_{2L} c_{1j} - g_{2R} c_{2j} = g (c_{1j} - c_{2j})$ are the vector and axial vector couplings of $\chi$ to the mediating gauge boson where j denotes the index corresponding to the mediating vector boson. These couplings are fixed from the gauge structure of the model. $c_{ij}$ are the elements of the $4 \times 4$ mass mixing matrix in the neutral gauge sector. The elements $c_{ij}$ are as the following \cite{32121},
	\begin{eqnarray}
	c_{11} = \cos\theta_W, c_{21} = -\sin\theta_W \tan\theta_W, c_{31} = \dfrac{-g_Y \sin\theta_W}{g_{1L}}, c_{41} = \dfrac{-g_Y \sin\theta_W}{g_{1R}} \nonumber \\
	c_{14} = \sin\theta_W, c_{24} = \sin\theta_W, c_{34} = \dfrac{\sqrt{\cos 2\theta_W}}{\sqrt{2}}, c_{44} = \dfrac{\sqrt{\cos 2\theta_W}}{\sqrt{2}} \nonumber \\
	c_{12} = -1.643\times 10^{-4}, c_{22} = 0.704, c_{32} = -0.707, c_{42} = 5.457 \times 10^{-2} \nonumber \\
	c_{13} = 2.255 \times 10^{-5}, c_{23} = -0.450, c_{33} = -0.386, c_{43} = 0.804 \nonumber
	\end{eqnarray} with $\theta_W$ being the Weinberg angle.
	
\begin{figure}[H]
		\begin{center}
			\includegraphics[width=8cm, height=3cm]{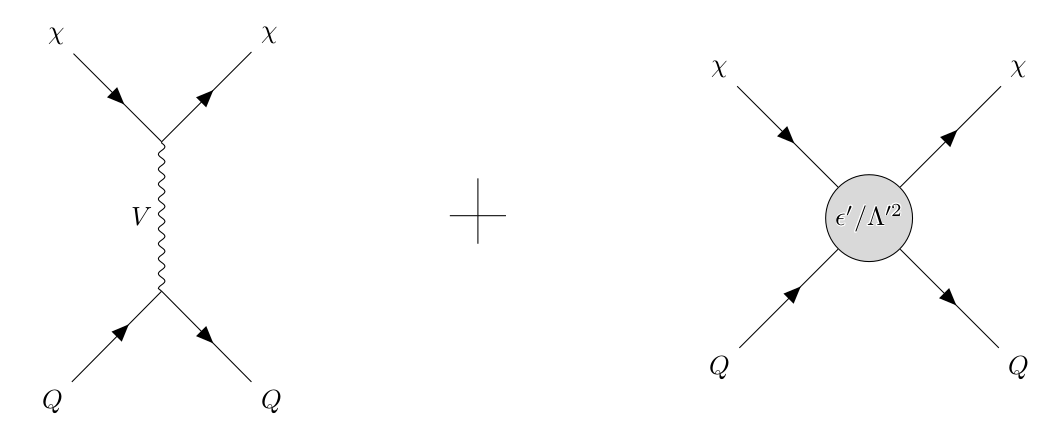}
		\end{center}
		\caption{Feynman diagrams (at quark level) responsible for the direct detection of DM ($\chi$) corresponding to the tree-level coupling present in 32121 model (left) and dim-6 effective operator (right). V is any heavy neutral gauge boson of 32121 model. The interference between these two diagrams will play a crucial role to the DM-nucleon scattering cross-section.}
		\label{effdiag}
	\end{figure}

\begin{figure}[h!]
		\begin{center}
			\includegraphics[width=8.5cm, height=6cm]{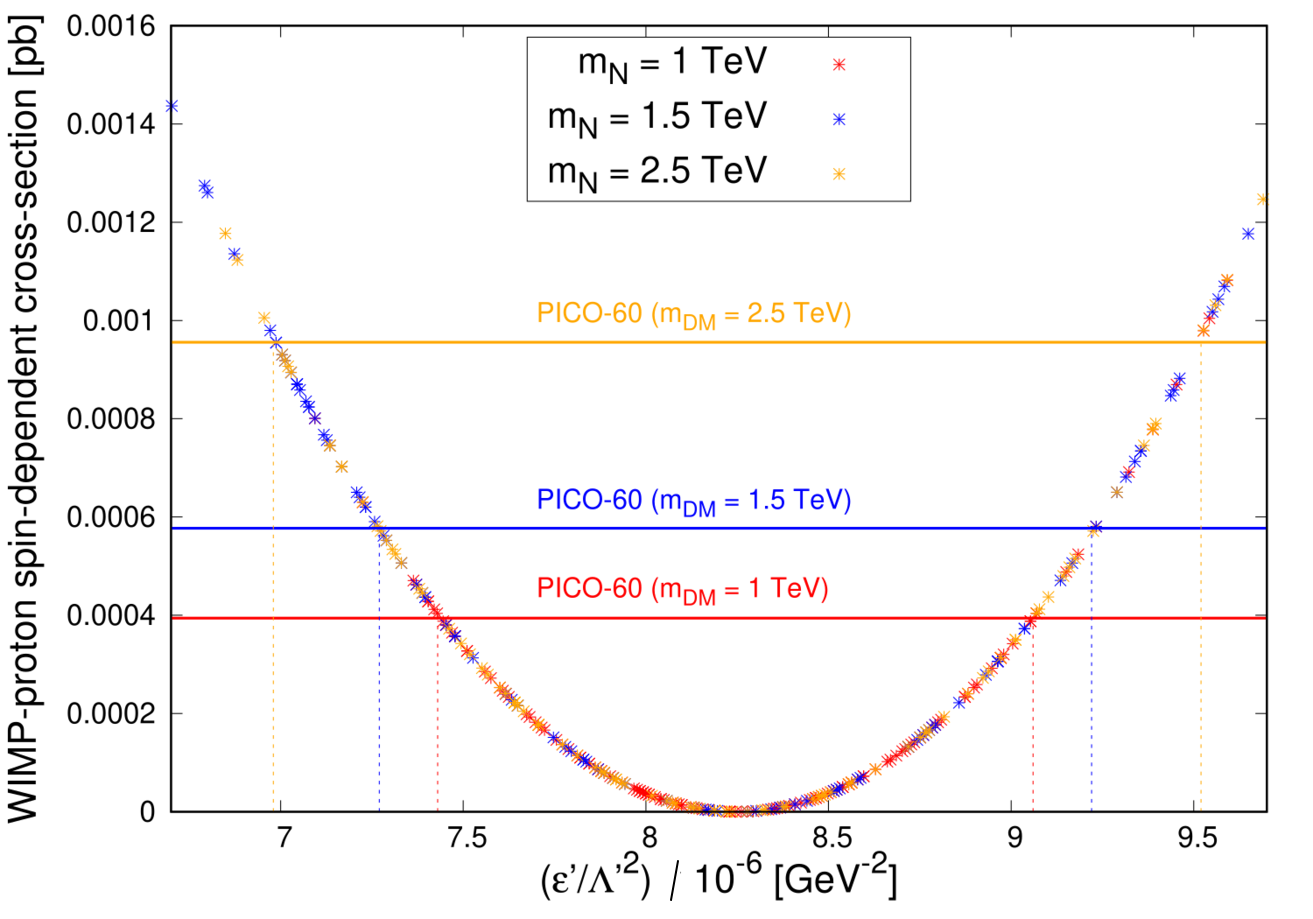}
			\includegraphics[width=8.5cm, height=6cm]{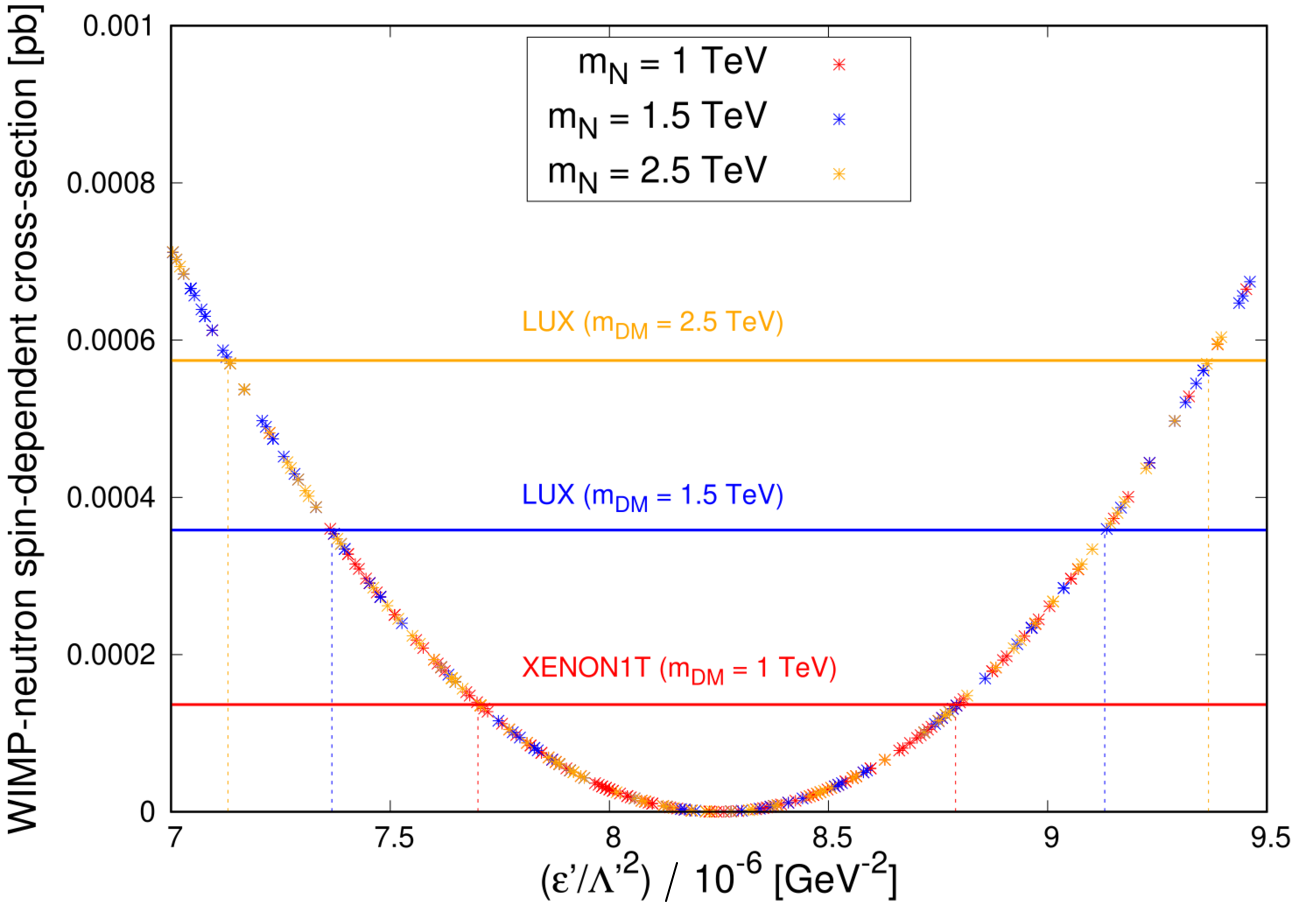}
		\end{center}
		\caption{Variation of $\sigma_{SD}$ with $\epsilon'/\Lambda'^2$ for WIMP-proton (left panel) and WIMP-neutron (right panel) elastic scattering. The red, blue and orange colors correspond to $m_{N}=1,1.5$ and $2.5$ TeV respectively. The solid lines represent the upper limits on $\sigma_{SD}$ from several direct detection experiments.}
		\label{eps}
	\end{figure}

	For $1$ TeV mass of $N$, $\sigma_{SI}=4.466\times 10^{-13}$ pb and $\sigma_{SD}=4.08\times 10^{-2}$ pb for proton-DM ($N$ in this case) scattering. We have observed that $\sigma_{SD}$ is order of magnitudes higher than $\sigma_{SI}$ and it also exceeds the upper limit of DM-nucleon scattering cross-section from PICO \cite{pico}. In order to rescue $N$ from such a conflict with experimental result, a dimension-6 four-fermion operator has been introduced (see Eq. \ref{NDDv_new}), with a hope that an appropriate choice of the effective coupling ($\epsilon'/\Lambda'^2$) could bring this spin-dependent cross section within the experimental limit. 
	\begin{equation}
	\label{NDDv_new}
	\dfrac{\epsilon'}{\Lambda'^2} Tr[(\bar{L}_B \gamma^\mu \tau_a L_B)]~(\bar{f}_L \gamma_\mu \tau_a f_L + L \leftrightarrow R)
	\end{equation}
	Such an interaction could have originated from the scattering of a pair $N$ off the quarks mediated by heavy Higgs boson belonging to $\bf 78$ representation of $E_6$ \footnote{Such an effective interaction could have been resulted from a scattering of say $N \bar N \rightarrow q \bar q$ mediated by a heavy  Higgs boson belonging to the {\bf 78} dimensional representation of $E_6$. Following \cite{slansky}, it is evident that {\bf 78} being an adjoint representation arising from the direct product of {\bf 27} with {$\bar {\rm \bf 27}$}. Mass scale of such a heavy Higgs boson can be in the range of 10s of TeV.  While considering DM phenomenology at TeV scale or below, we can safely integrate out such heavy degrees of freedom leading to such effective dimension-6 interactions.}. Here, $\Lambda'$ is a heavy mass scale probably related to the mass of the Higgs boson in consideration and $\sqrt{\epsilon'}$ is proportional to the relevant gauge coupling. $\tau_a$s are the Pauli matrices ($a = 1,2,3$), $f$ is any SM fermion multiplet. 
	Fig. \ref{effdiag} shows the two Feynman diagrams at quark level who will be responsible for the DM-nucleon scattering. The left diagram is solely controlled by the gauge coupling constants of 32121 model and the right diagram represents the dim-6 effective operator. For the spin-dependent scenario, the contribution from the left diagram will be too high. But there will be a significant contribution from the interference between the two diagrams which will help to compensate the effect of the left diagram and bring down the total DM-nucleon scattering cross-section below the experimental limit.
	One can see, while calculating the scattering rate, choice of a negative $\epsilon'$ can in turn reduce the whole spin-dependent scattering cross-section and satisfy the upper limit on $\sigma_{SD}$ arising from the present direct detection experiments.
	In the following work, we have implemented this model in \texttt{FeynRules} \cite{feynrules} and have used the package \texttt{micrOMEGAs5.2} \cite{micromegas} to calculate the direct detection cross sections and relic abundances. In Fig. \ref{eps}, the variation of $\sigma_{SD}$ for DM-nucleon scattering with $\epsilon'/\Lambda'^2$ has been shown for three values of masses of $N$.
	One can see from Fig. \ref{eps}, the allowed range of $\epsilon'/\Lambda'^2$ is $7.43-9.06 \times 10^{-6}$ GeV$^{-2}$ ($7.7-8.79 \times 10^{-6}$ GeV$^{-2}$) for DM-proton (neutron) scattering. for an $N$ with mass of 1 TeV. The greater the mass of $N$, a wider range of $\epsilon'/\Lambda'^2$ is allowed by the data from the experiments like XENON \cite{direct-detection}, LUX \cite {lux} or PICO-60 \cite{pico}. The range of values for $\epsilon'/\Lambda'^2$ allowed from WIMP-neutron scattering are more stringent than the range of values obtained from WIMP-proton scattering (see Fig. \ref{eps}).

Allowed values of the effective coupling $\frac{\epsilon '}{\Lambda '^2}$ at low energy, imply a relatively large Yukawa coupling of the order of $8$ between the SM/exotic fermion $N$ with Heavy Higgs bosons of the theory. 
The exact value of such Yukawa couplings can be calculated using RG 
	running of the couplings with some assumptions (about the boundary condition) of the couplings at a high scale in the framework of $E_6$. However, this is beyond the scope of the present work and we take this value of the effective coupling as an experimental constraint on such a model.

	\vspace{5mm}
	
	$L_S$ could well be a candidate of being a relic particle along with $N$. Table \ref{table1} shows $L_S$ has exactly equal and opposite $U(1)_{L,R}$ charges which in turn prohibits its coupling to the SM $Z$-boson. It has axial-vector couplings to heavy gauge bosons $A'$ and $Z'$. It also couples to scalars via a dimension-5 operator mentioned in Eq. \ref{LRYukawa5}.
	Elastic scattering of $L_S$ with the nucleons are mediated either by scalars or by heavy gauge bosons (see Fig. \ref{NDDdiag}) and can be parametrized via the operators similar to those in Eqs. \ref{NDDs}, \ref{NDDv}.
	These interactions are very similar to $N$ apart from the absence of vector like coupling, $g_{\chi V}$ (see Eq. \ref{NDDv}) due to the Majorana nature of $L_S$ and contains an extra factor of $1 \over 2$ in each case.
	The definition of the couplings ($g$ and $\mathcal{C}$) are similar as in cases of Eqs. \ref{NDDs} and \ref{NDDv}.
	
	In all practical purpose, scalar mediated diagrams do not contribute to the elastic scattering cross-section for the reasons mentioned before. Primarily, scalar couplings to the SM quarks are proportional to fermions masses while the scalar coupling to the a pair of $L_S$ is proportional to $v_s \over \Lambda$ which could be a number order of one. However, overall rate of this scattering is small due to the small admixture of $\phi_S$ in the SM Higgs boson \cite{32121}.	 	
	Thus one is left with the contributions from vector mediated processes following Eq. \ref{NDDv} with $g_{\chi V}=0$ to obtain spin-dependent scattering cross-section of DM over nucleons. We shall have a similar expression like Eq. \ref{Nsigma_SD} as $\sigma_{SD}$. 
	
	Although the rate of gauge mediated process is proportional to gauge couplings, in presence of heavy $A'$ and $Z'$ as the mediator, finally it results into a small $\sigma_{SD}$ for $L_S$ ($\sim 10^{-10}$ pb over a wide range $L_S$ mass varying from 1 to 4 TeV), well below the experimental upper limits by LUX or PICO.
	
	Before we go into the next section to discuss the issue of relic density, few comments about our results are in order. 
	We have already pointed out that direct detection limits can only be satisfied for $N$, with a non-zero $\epsilon'$ and we have identified a range of values of ${\epsilon'} \over {\Lambda'} ^2$ by comparing the direct detection cross-section with experimental data. So instead of looking for low mass $N$, we will concentrate in the case when a non-zero value of ${\epsilon'} \over {\Lambda'}^2$ has been assumed. Such an effective operator along with gauge interactions will contribute to the annihilation of $N$ to a pair of SM fermions. When such allowed (by direct detection) values of ${\epsilon'} \over {\Lambda'} ^2$ have been used to estimate the annihilation cross-section of $N$ the thermal averaged cross-section becomes too large making the relic density too small ($\sim 10^{-4}$).

	\subsection{Estimation of Relic Density}
	
	 \underline{\bf Case of $y_{BB} =0$ :} Barring any interaction between $N$ and $L_S$ (setting $y_{BB} = 0$), the heavier between these two does not decay to anything else and practically becomes stable. Thus both $N$ and $L_S$ can contribute to relic density. We will investigate whether this scenario could yield sufficient amount of relic which is consistent with experimental data.
		
	In the early Universe, when the temperature of the Universe $T \gg m_{\chi}$, $\chi$ (the particle we consider as dark matter) was abundant and in thermal equilibrium with the SM particles. As the Universe cools down and $T < m_{\chi}$, $\chi$ decouples from the thermal bath. When the annihilation rate of $\chi$ is equal to the Hubble expansion rate, it freezes out and the abundance of $\chi$ becomes constant. The evolution (with time/temperature) of the number density of the relic particle of our interest can be estimated by solving Boltzmann equation \cite{srednicki}.

	In a two-component dark matter scenario like ours, let $\chi_1$ and $\chi_2$ are the DM particles of mass $m_{\chi_1}$ and $m_{\chi_2}$ respectively. $n_1,n_2$ are the corresponding number densities of the relic particles. One can identify $\chi_1$ and $\chi_2$ with $L_S$ and $N$ respectively. Boltzmann equations are as the following \cite{boltzman},
	
	\begin{eqnarray}
	\dfrac{dn_1}{dt} = &-& 3 H n_1 - \langle \sigma_1 v \rangle (n_1^2-{n_1^{eq}}^2) - \langle \sigma_{11\rightarrow 22} v \rangle (n_1^2-\dfrac{{n_1^{eq}}^2}{{n_2^{eq}}^2} n_2^2) \Theta(m_{\chi_1}-m_{\chi_2}) \nonumber \\
	&-& \langle \sigma_{22\rightarrow 11} v \rangle (n_2^2-\dfrac{{n_2^{eq}}^2}{{n_1^{eq}}^2} n_1^2) \Theta(m_{\chi_2}-m_{\chi_1}) - \sum_{j, j \neq 1} \langle \sigma v \rangle_{1j} (n_1 n_j -n_1^{eq} n_j^{eq}) \\
	\dfrac{dn_2}{dt} = &-& 3 H n_2 - \langle \sigma_2 v \rangle (n_2^2-{n_2^{eq}}^2) - \langle \sigma_{22\rightarrow 11} v \rangle (n_2^2-\dfrac{{n_2^{eq}}^2}{{n_1^{eq}}^2} n_1^2) \Theta(m_{\chi_2}-m_{\chi_1}) \nonumber	\\
	&-& \langle \sigma_{11\rightarrow 22} v \rangle (n_1^2-\dfrac{{n_1^{eq}}^2}{{n_2^{eq}}^2} n_2^2) \Theta(m_{\chi_1}-m_{\chi_2}) - \sum_{j, j \neq 2} \langle \sigma v \rangle_{12} (n_j n_2 -n_j^{eq} n_2^{eq})
	\end{eqnarray}
	where, $\langle \sigma_{1,2}v \rangle$ are the thermally averaged annihilation cross-section of a pair of $\chi_1$ or $\chi_2$ to lighter SM particles ($\chi_{1,2}~ \chi_{1,2} \rightarrow f f$), $\Theta$ is the Heaviside function. $\langle \sigma_{ii\rightarrow jj} v \rangle$ corresponds to the DM-DM conversion processes ($\chi_{1(2)}~ \chi_{1(2)} \rightarrow \chi_{2(1)}~ \chi_{2(1)}$) when kinematically allowed. The last term represents co-annihilation of two DM candidates ($\chi_{1}~ \chi_{2} \rightarrow f f$) or co-annihilation of one of the relic particles with any other which is close to mass with the relic particle (in our case, this could be identified with $E^\pm$). Here, we would like to remind that throughout our analysis, $E$ remains heavier in mass than $N$. However, for $y_{BB} =0$, $E$ and $N$ becomes mass degenerate at tree level. But as we mentioned before, $E$ still can be heavier due to extra radiative corrections its receives due to electro-magnetic interactions. Without going into the details of such corrections we will assume $m_E - m_N = 1 ~\rm GeV$ throughout the analysis when $y_{BB} = 0$.

	N or $L_S$ in general can annihilate to a pair of SM fermions, gauge bosons (neutral and charged) and scalars. For example, annihilation of $\chi$ to a pair of SM fermions via gauge bosons can be driven via following interactions.
	\begin{eqnarray}
	\mathcal{L} \subset m_V^2 V^\mu V_\mu + \bar{\chi} \left(i \gamma_{\mu} \partial^{\mu} - m_{\chi}\right) \chi + \bar{f} \left(i \gamma_{\mu} \partial^{\mu} - m_{f}\right) f \nonumber \\ 
	+ \bar{f} \left(g_V \gamma_{\mu} + g_A \gamma_{\mu} \gamma^5 \right) f V^\mu + n \bar{\chi} \left(g_{\chi V} \gamma_{\mu} + g_{\chi A} \gamma_{\mu} \gamma^5 \right) \chi V^\mu 
	\end{eqnarray}
	where $g_V$ ($g_{\chi V}$) and $g_A$ ($g_{\chi A}$) are vector and axial-vector like couplings between SM-fermions (DM candidate $\chi$) and mediating gauge boson respectively. These couplings are fixed from the gauge structure of the model. It is to be noted that $n=1$ for $N$ (Dirac fermion) and $1/2$ for $L_S$ (Majorana fermion). Further for $L_S$, $g_{\chi V}=0$. $m_f, m_{\chi}, m_V$ are the masses for final state fermions, dark matter and mediator respectively.
	
	The thermally averaged cross-section, $\langle \sigma v \rangle$ can be expanded in powers of $v^2$ or $x (=T/m_{\chi})$ (as $v \sim \sqrt{T}$) like,
	\begin{equation}
	\langle \sigma v \rangle = a_0 + b_0 x + c_0 {x^2} + ...
	\end{equation}
	$a_0,b_0,c_0...$ are model dependent constant parameters depending on the couplings and masses of the particles taking part in the annihilation process. 
	As for example, expressions of $a_0$ and $b_0$ for $\chi \chi \rightarrow f \bar{f}$ can be expressed as \cite{srednicki},
	\begin{eqnarray}
	a_0 &=& \dfrac{N_c n^2 \beta}{2 \pi m_V^4 (m_V^2-4m_{\chi}^2)^2}\left[g_A^2 \left(g_{\chi A}^2 m_f^2 (m_V^2-4m_{\chi}^2)^2 + 2 g_{\chi V}^2 m_V^4 (m_{\chi}^2-m_f^2) \right) + g_V^2 g_{\chi V}^2 (m_f^2 + 2m_{\chi}^2)\right] \\
	b_0 &=& \dfrac{3}{4} \dfrac{m_f^2}{m_{\chi}^2-m_f^2} a_0 + \dfrac{N_c n^2 \beta}{48 \pi m_{\chi}^2 m_V^4 (m_V^2-4m_{\chi}^2)^2} [ 12 g_A^2 m_{\chi}^2 m_V^4 \left\lbrace 4 m_f^2 g_{\chi V}^2 - 5 m_f^2 g_{\chi A}^2 + m_{\chi}^2 \left(2g_{\chi A}^2-g_{\chi V}^2 \right) \right\rbrace \nonumber \\
	&+& 144 g_A^2 g_{\chi A}^2 m_{\chi}^4 m_f^2 m_V^2 + 6 g_V^2 m_{\chi}^2 m_V^4 \left\lbrace 2g_{\chi A}^2 m_f^2 - 3 g_{\chi V}^2 m_f^2 + m_{\chi}^2 \left(4g_{\chi A}^2 - g_{\chi V}^2 \right) \right\rbrace]
	\end{eqnarray}
	$N_c$ is the color factor, $N_c=1$ for leptons and $3$ for quarks. $\beta = \sqrt{1-m_f^2/m_{\chi}^2}$. 
	
	An approximate solution of the Boltzmann equation gives us the present mass density or relic density as the following \cite{subhaditya},
	\begin{eqnarray}
	\Omega_{L_S} h^2 \simeq \dfrac{0.1}{{\langle \sigma v \rangle_{eff}}_{L_S}} ~pb \nonumber \\
	\Omega_{N} h^2 \simeq \dfrac{0.1}{{\langle \sigma v \rangle_{eff}}_{N}} ~pb \nonumber \\
	\Omega_{\chi} h^2 = \Omega_{L_S} h^2 + \Omega_N h^2
	\end{eqnarray}
	$\langle \sigma v \rangle_{eff}$ is the total effective annihilation cross-section of the corresponding relic particle.
	
\begin{figure}[h!]
		\begin{center}
			\includegraphics[width=8cm, height=6cm]{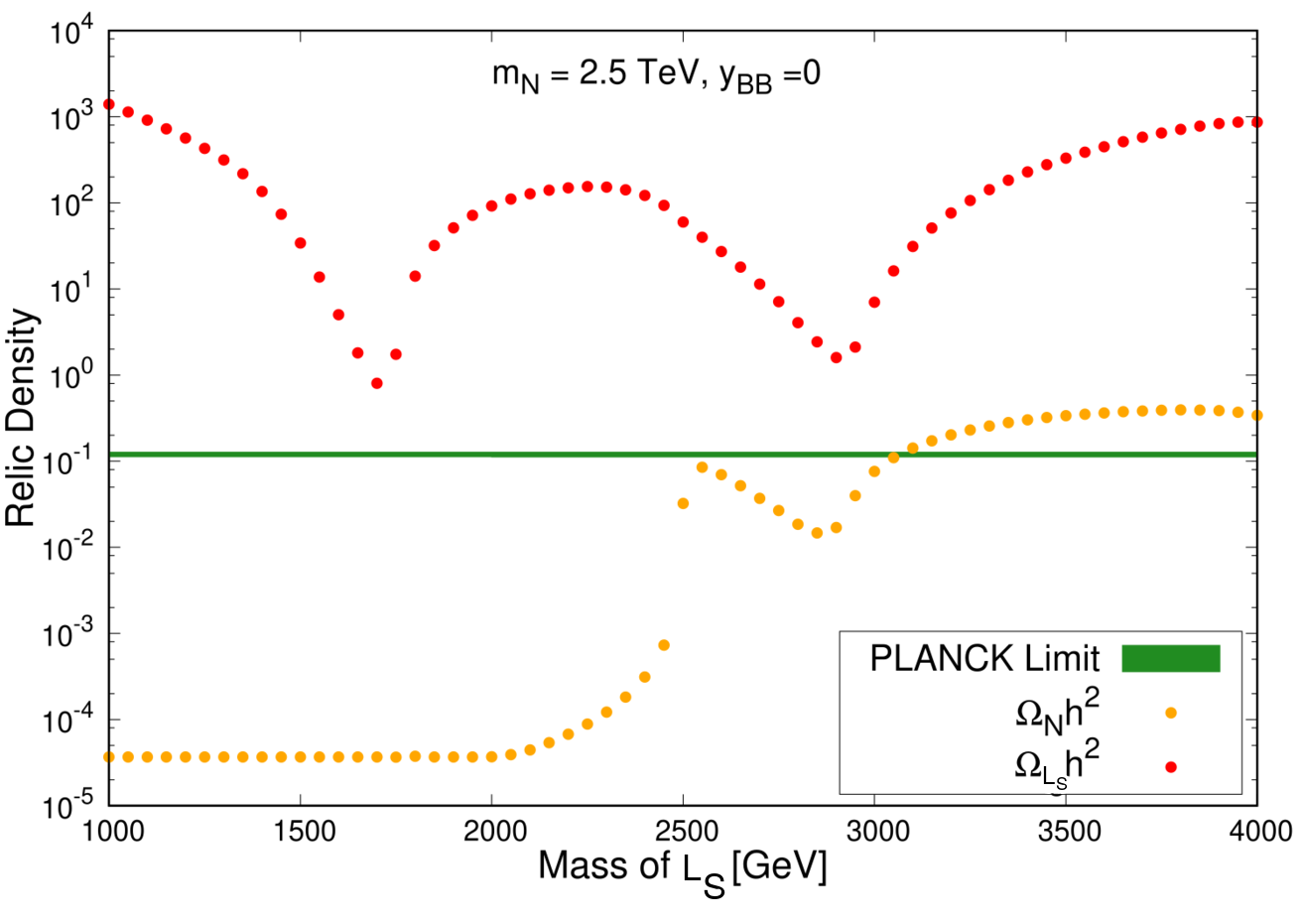}
			\includegraphics[width=9cm, height=6cm]{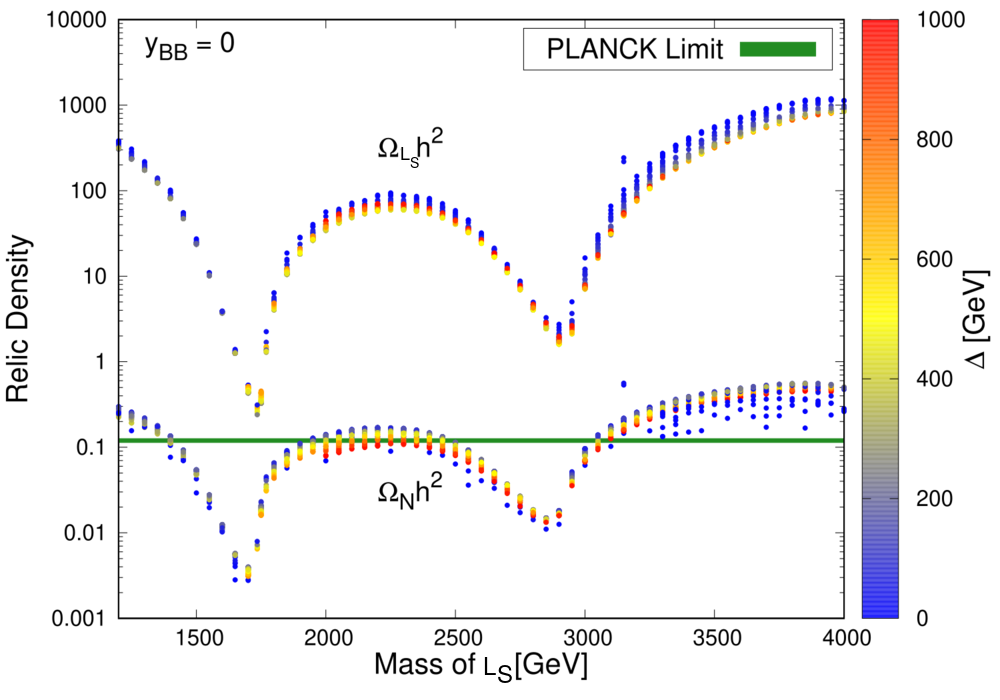}
		\end{center}
		\caption{Relic density of each DM vs mass of $L_S$ plot with fixed values of the heavy neutral gauge bosons for $m_N = 2.5$ TeV (left panel) and for $m_{L_S} > m_N$ with varying $\Delta$ (right panel). In both cases $y_{BB} = 0$.}
		\label{LSRelic}
	\end{figure}

	Annihilation channels available to $L_S$ are, to a pair of SM particle via $A'$ and $Z'$ exchange in s-channel.
	$L_S$ being an $SU(2)_{L,R}$ singlet its annihilation is controlled by the $U(1)$ gauge couplings and $U(1)_{L,R}$ quantum numbers.
	The variation of relic density $\Omega_{L_S} h^2$ with its mass has been presented in Fig. \ref{LSRelic}. 
	Setting the masses of $A'$ and $Z'$ to their lower limits results into an annihilation cross-section finally leading to an over-abundance of relic particle at present epoch. 
	The mass difference between $N$ and $L_S$ ($\Delta=\vert m_N - m_{L_S} \vert$) will play a crucial role in determining the relic. In the following we discuss this issue in detail. 
	
	On the other hand, the annihilation cross-section of $N$ is much higher than that of $L_S$. This can be accounted by the $SU(2)$ gauge interactions of $N$ as well as a dimension-6 four-fermi interaction (proportional to $\epsilon'$) involving $N$ and SM quarks. So when $m_{L_S}$ approaches $m_N$, (within $10-15\%$ of $m_{N}$) $L_S L_S \rightarrow N N$ annihilation will increase hence increasing the relic density for $N$. This is evident from the left panel of Fig. \ref{LSRelic}. The right panel of Fig. \ref{LSRelic} explains the variation of relic density of each DM candidate with the mass difference $\Delta$ where $m_{L_S}$ has been set to higher value than $m_N$ so that throughout the region the DM conversion ($L_S L_S \rightarrow N N$) can assist to reduce $\Omega_{L_S} h^2$ and increase $\Omega_{N} h^2$. Inspite of that one can observe even a sufficient mass difference $\Delta$ can lower the relic density of $L_S$ but does not succeed to satisfy the PLANCK limit. 
	The sudden decrement of relic density both for $L_S$ and $N$ around $m_{L_S} = 3~\rm TeV$ is due to the enhanced annihilation rate to the SM particles via $Z'$ resonance. Similar enhancement of annihilation to the SM is also observed for $L_S$ when the same is driven by $A'$ resonance for $m_{L_S} \simeq 1.7~ \rm TeV$. Even, near the $A'$ and $Z'$ resonances, annihilation cross-section is not big enough to produce the required amount of relic as measured by PLANCK. For higher values of mediating gauge boson masses, annihilation cross-section will further reduce and it would be more difficult to satisfy the relic abundance limit.
	
	Few comments about the input parameters used in our analysis are in order. In general, the Higgs sector of LRS model is parametrised by a large number of parameters. The relic particles in our case couples to the SM particles via the SM like Higgs boson, the $SU(2)$ singlet Higgs boson and the gauge bosons.  Their couplings to the gauge bosons are model independent and completely determined by their gauge quantum numbers. However, in general, DM couplings to Higgs bosons depend on the Yukawa couplings along with the Higgs sector parameters. Elements of the mixing matrix which connects the physical scalars with gauge eigenstate scalars, are determined from the scalar sector parameters. 
	However, the experimentally measured signal strengths of the SM Higgs restrict the mixing of SM like Higgs with any other scalar to be tiny, resulting into a practically near diagonal scalar mixing matrix \cite{32121}. This facilitates us to treat the physical masses of the scalars as free parameters instead of using plethora of scalar sector parameters as input variables. Hence, we can encapsulate the interaction of relic particles without considering the details of several scalar sector parameters. Let us now consider the case of interaction of relic particles with SM fermions and gauge bosons via the SM Higgs. Couplings of SM Higgs to the SM fermions are proportional to the SM fermions masses and couplings to the gauge bosons depend on the gauge coupling constants. Consequently such couplings are model independent. Such annihilation rates thus depend on the Yukawa coupling $y_{LB}$ or $y_{LS}$. 
	Now, the relic particles also can annihilate to heavy gauge bosons via the mediation of singlet Higgs. Singlet Higgs couplings to gauge boson again is determined by the gauge coupling constants only. While the DM-singlet Higgs coupling is proportional to $y_{LS}$ or $y_{LB}$. Masses of the relic particles are proportional to these Yukawa couplings and are free parameters of our analysis. However, as the heavy gauge bosons have masses of the order of few TeV, such annihilation rates are already small.

	However, the results presented in Fig. \ref{LSRelic} poses us with a problem. First of all, the relic density calculated in such a scenario, in which both $N$ and $L_S$ are contributing, is orders of magnitude higher than the measured value of the same from PLANCK experiment, and thus is ruled out. To overcome such a situation, a tri-linear interaction involving $L_S$, $N$ and $h^0$ (proportional to $y_{BB}$) would help us and now we will discuss this issue in the following. 	\\\\

\begin{figure}[H]
		\begin{center}
			\includegraphics[width=10cm, height=4cm]{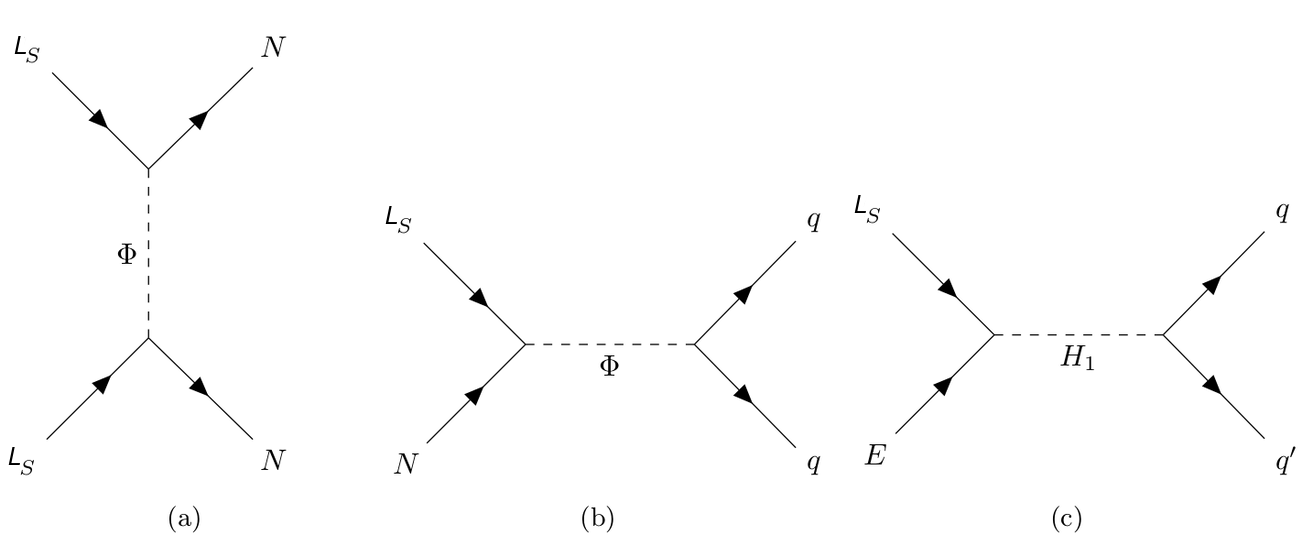}
		\end{center}
		\caption{The DM annihilation and co-annihilation channels opened up by a non-zero $y_{BB}$}
		\label{CoAnnDiag}
	\end{figure}

	\underline{\bf Case of $y_{BB} \neq 0$ :} A non-zero $y_{BB}$ (see Eq. \ref{LRYukawa}) will open up the channels like Fig. \ref{CoAnnDiag}. The diagram (a) explains the annihilation of a pair of $L_S$ to a pair of $N$. The diagrams (b) and (c) represent co-annihilations between $N/E$ and $L_S$ where $q,q'$ are SM particles. Diagram (c) is only important in case of third-generation of quarks ($t,b$) as the couplings of $H_1^\pm$ to the other SM quarks and leptons are negligibly small \cite{32121}. Here, $\Phi$ represents any one of the neutral scalars, $h^0$, $h_2^0$, $\xi_2^0$ and $H_S^0$. All such processes will reduce the relic of $L_S$ and $N$.

	One may assume $N$ being heavier than $L_S$ and treat the later with relic particle. However this scenario will produce a relic density much higher than the experimentally measured value. Such a result can be accounted by low interaction rate of $L_S$ along with the decay of $N$ to $L_S$ driven by non-zero $y_{BB}$ couplings. The complementary scenario in which $N$ being lighter than $L_S$ can also be considered. This scenario, on the contrary, produce relic density much smaller than the experimentally measured value, which is an outcome of high annihilation rate of $N$ to SM particles. So we have no other option than to consider a situation when both $N$ and $L_S$ are relic particle. This could be realised only if we can prohibit the decay of $N$ or $L_S$ via $y_{BB}$ couplings or we may choose the mass difference between the DM candidates too small so that the decay lifetime will be comparable with that of the Universe. A very large lifetime ($\sim 10^{22} s.$) of the aforementioned decay could have been possible with a very tiny mass difference between $N$ and $L_S$ or with an absymaly small value of $y_{BB}$. From the discussion in the previous section (case of $y_{BB} = 0$), it is evident that the second choice may not solve the problem of overabundance of relic particles. While choosing a very small mass difference between two TeV scale particles will practically mean degeneracy between them.
	Eq. \ref{Nmass} reveals that there will be a small mass difference between $N$ and $L_S$ generated via the Yukawa interaction. Apart from that, $N$ and $L_S$ can have different radiative contributions to their masses as their gauge transformation properties are different. That could compensate the mass difference generated by Yukawa interaction and make them degenerate in mass. So in the rest of the analysis apart from assuming a non-zero interaction among $N$ and $L_S$, a complete mass degeneracy between these two particles (i.e., $m_N=m_{L_S}$) have also been assumed. A possible mechanism for mass degenerate relic particles has been discussed at the end of section \ref{subsec21}.
	\vspace{5mm}

	A larger mass splitting between $E$ to $N$ gives rise to a higher rate of non-thermal production \cite{non-thermal} of $N$ (via the decay $E \rightarrow N W (W')$). This in turn could increase the relic abundance of $N$. However, following an earlier discussion (in section \ref{subsec21}) it has been emphasised that $E$ and $N$ can have mass separation as large as $\mathcal{O}(10 \mbox{~GeV})$ for the ranges of values of $y_{BB}$ and $m_N$ used in our analysis. Such a, small mass difference, would keep the rate of non-thermal production negligibly small in comparison to the thermal production mechanism.	

	For the present analysis with $N$ and $L_S$ as mass degenerate dark matter particles we have fixed the value of $\epsilon'/\Lambda'^2 ~(8.16\times 10^{-6}$ GeV$^{-2})$. It has been mentioned that the annihilation of $N$ is sensitive to this parameter. This particular value of $\epsilon'/\Lambda'^2 ~(8.16\times 10^{-6}$ GeV$^{-2})$ is chosen so that for any value of $m_N$ the direct detection limit could be satisfied which is evident from Fig. \ref{eps}.
	
	\begin{figure}[h!]
		\begin{center}
			\includegraphics[width=10cm, height=6cm]{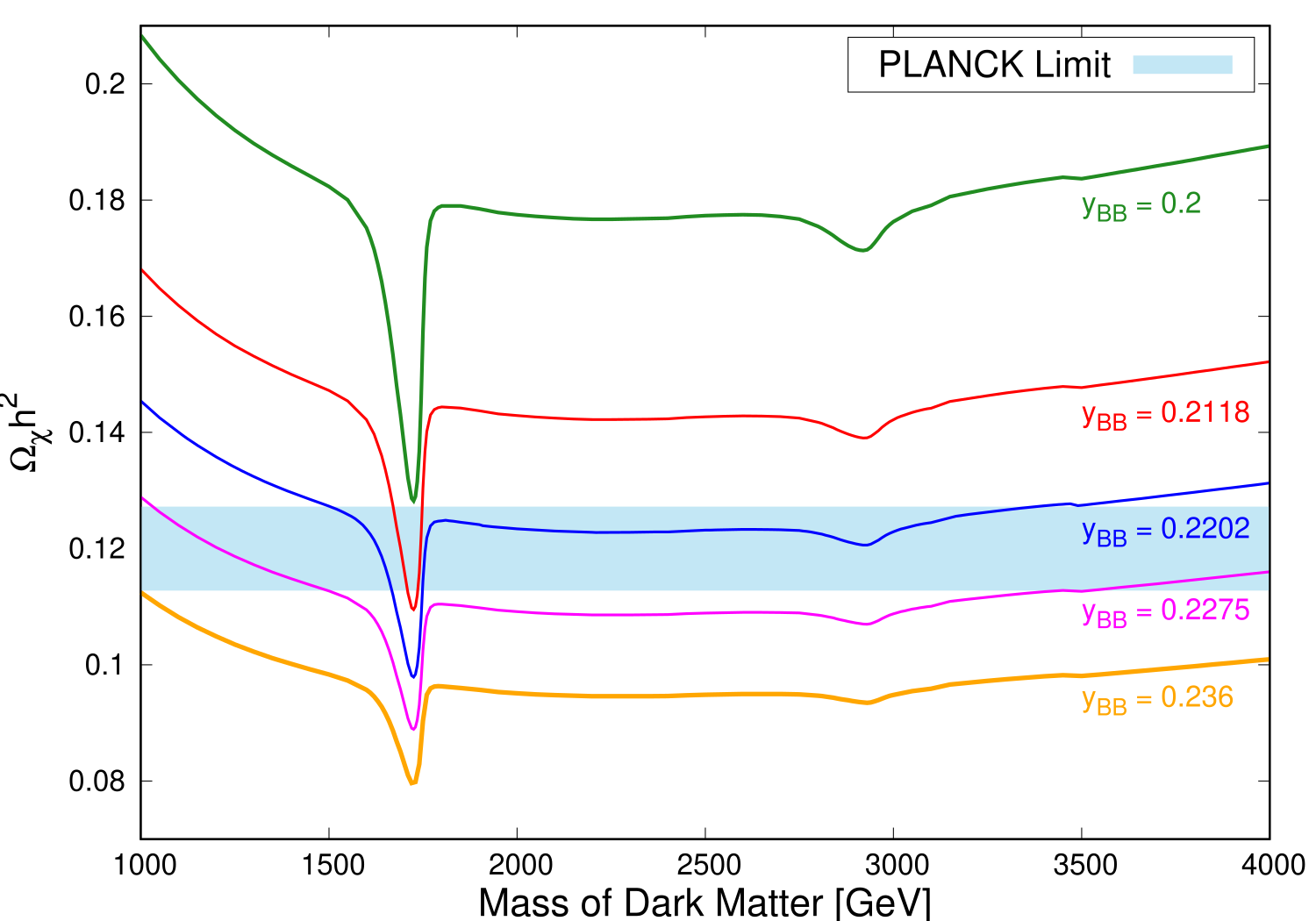}
		\end{center}
		\caption{Variation of total relic density over a wide range of mass of dark matter for different $y_{BB}$. The masses of the heavy neutral gauge bosons are fixed at their lower limits, $m_{A'}=3.5$ TeV and $m_{Z'}=5.89$ TeV. Mass degeneracy between $N$ and $L_S$ has been assumed. }
		\label{ybbrelic}
	\end{figure}

	Now we are ready to discuss the results obtained from our analysis treating both $L_S$ and $N$ as the dark matter. Relic density as a function of relic particle mass has been presented in Fig. \ref{ybbrelic} for different values of $y_{BB}$. It is evident from the plots that resultant relic density is very sensitive to $y_{BB}$ and unless the value of this particular Yukawa coupling is greater than a critical value (0.2) one ends up with scenario with over-abundance of relic particle. A sharp fall of relic density around $m_{L_S} \simeq 1.7 \rm ~TeV$ is due to the resonant annihilation of relic particles via $A'$ (of mass $3.5 \rm ~TeV$) and a not so prominent fall of the relic density around $m_{L_S} \simeq 2.9 \rm ~TeV$ is due to the annihilation of the relic via a $Z'$ exchange. It is to be noted that rate of annihilation via Higgs exchange is dominant over the rate of annihilation of relic particles via gauge boson exchange. This has been manifested in the the gradual decrease of the sharpness of the resonant peaks for higher and higher values of $y_{BB}$.

	In this whole analysis, the masses of the gauge bosons have been set at their lowest limits, $m_{A'} = 3.5$ TeV, $m_{Z'} = 5.89$ TeV, $m_{W'} = 4.8$ TeV. For the case where $y_{BB} = 0$, we will not be interested in such investigation as the larger mass of the gauge bosons will lead to an over-abundant scenario. In general, increasing the gauge bosons masses will reduce the annihilation cross-section of relic particles leading to increase of relic abundance. However, a higher value of $y_{BB}$ will increase the co-annihilation cross-section thus bring down the relic abundance in the desired range. In such a case, we obtain a more restricted range of $y_{BB}$. In Table \ref{table2}, the allowed ranges of $y_{BB}$ have been noted for different masses of the gauge bosons, scanned over the mass range of DM as shown in Fig. \ref{ybbrelic} is presented. It is to be noted that throughout our analysis, the mass difference of E and N has been set to be 1 GeV which could be generated via electromagnetic radiative corrections.
		\begin{table}[H]
			\centering
			\begin{tabular}{|c|c|c|}
				\hline \hline	
				$m_{A'}$ [TeV] & $m_{Z'}$ [TeV] & Allowed $y_{BB}$  \\ [1.0ex]
				\hline \hline				
				3.5 & 5.89 & $0.2 < y_{BB} < 0.236$ \\ [1.0ex]
				4 & 6.2 & $0.215 < y_{BB} < 0.235$ \\ [1.0ex]
				5 & 6.2 & $0.216 < y_{BB} < 0.235$ \\ [1.0ex]
				6 & 7 & $0.218 < y_{BB} < 0.235$ \\ [1.0ex]
				\hline \hline
			\end{tabular}
			\caption{The allowed range of $y_{BB}$ for different masses of gauge bosons is presented. The range of $y_{BB}$ gets more and more restricted while increasing the masses of the gauge bosons.}
			\label{table3}
	\end{table}

\begin{figure}[H]
		\begin{center}
			\includegraphics[width=10cm, height=6cm]{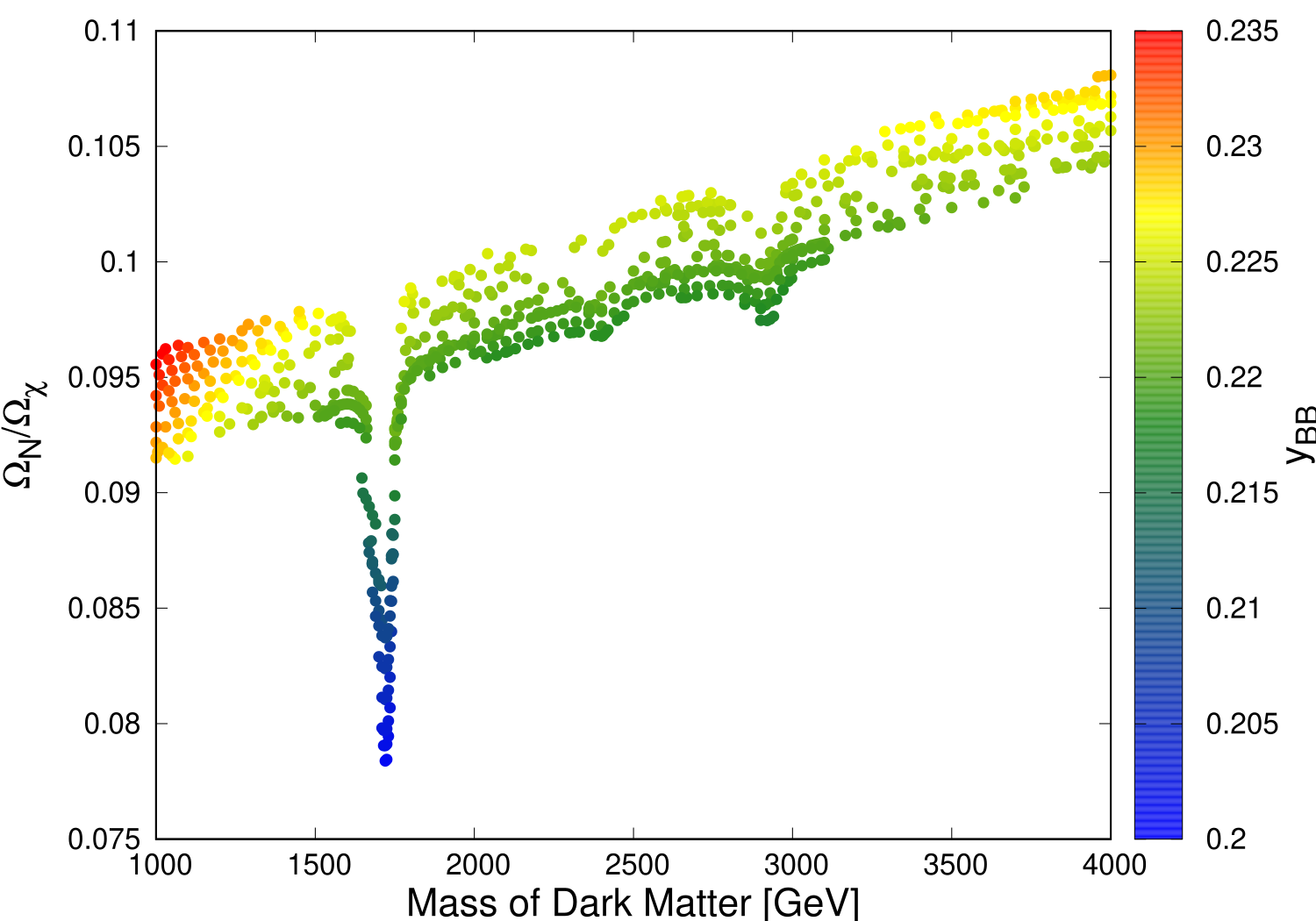}
		\end{center}
		\caption{Variation of the relative abundance of $N$ ($\Omega_{N}/\Omega_{\chi}$) with respect to mass of $N$ and $y_{BB}$. Relative abundance of $L_S$ can be understood as $1-\Omega_{N}/\Omega_{\chi}$. $m_N = m_{L_S}$ has been assumed.}
		\label{omegarelative}
	\end{figure}

	In Fig. \ref{omegarelative}, we present the relative abundance of $N$ (i.e., $\Omega_{N}/\Omega_{\chi}$) with the mass of $N$ for different values of $y_{BB}$. Over a wide range of mass of $N$, $L_S$ contributes dominantly to the relic density. This is mainly due to its lower interaction rate compared to $N$. The dips near $m_{DM} \simeq m_{A'}/2$ and $m_{Z'}/2$ also explain the higher annihilation rate of $N$ near the resonances. From Fig. \ref{omegarelative} (for $\epsilon'/\Lambda'^2=8.16\times 10^{-6}$ GeV$^{-2}$), one can see that $N$ can at most contribute 8-10.5$\%$ to $\Omega_{\chi}$. 
	In this context, it needs to be mentioned that, the value of the relative abundance of $N$, increases with decreasing $\epsilon'/\Lambda'^2$ (as a lower value of $\epsilon'/\Lambda'^2$ will decrease the annihilation rate of $N$). With a higher mass of $N$, the allowed range of $\epsilon'/\Lambda'^2$ becomes broader i.e., a lower value of this effective coupling becomes allowed (see Fig. \ref{eps}). Thus we can obtain a higher relative abundance of $N$ for a heavier $N$ which is evident from Fig. \ref{omegarelative}.
		
	Before we close this section, let us briefly comment on an effect which is relevant in our case, namely, the Sommerfeld enhancement of annihilation rate of DM due to self interaction. Both the DM candidates of our interest, have gauge interactions and effective interaction arising from exchange of massive gauge bosons would enhance the annihilation rate to great extent depending on the masses of DM, exchanged gauge boson and interaction strength. Depending on the velocity of the DM particle, which in our case can be extremely small, the annihilation rate can be enhanced by a factor of $10^5$ for some specific values of DM mass \cite{sommerfeld}. In turn, for such DM masses, relic abundance will be reduced by the same factor. We have to choose a non-zero $y_{BB}$, as putting off this interaction will yield overabundance of $L_S$, which is unacceptable. However, inclusion of Sommerfeld enhancement effect may thus bring down the relic density below the experimental limit for such case also. However, we have not included this effect in our analysis.

	 \subsection{Direct detection rate in case of $y_{BB} \neq 0$ :} 
\begin{figure}[H]
		\begin{center}
			\includegraphics[width=10cm, height=7cm]{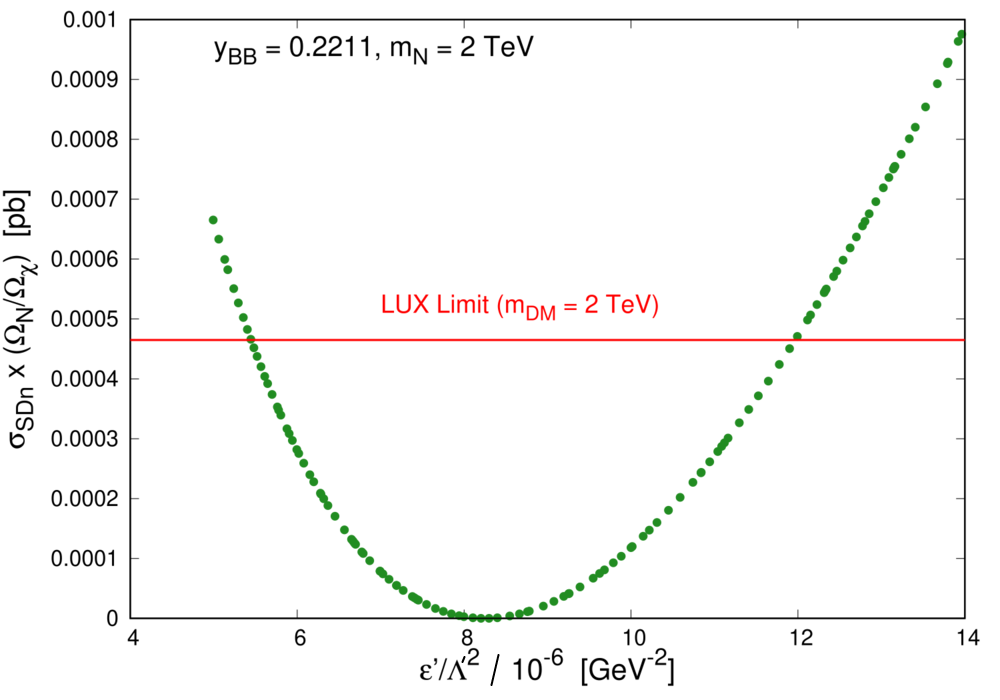}
		\end{center}
		\caption{Variation of spin-dependent scattering cross-section for a 2 TeV $N$ mass with the coefficient $\epsilon'/\Lambda'^2$. The green solid points represents the case for $y_{BB}=0.2211$ allowed by the PLANCK limit.}
		\label{new_eps}
	\end{figure}

In case of a two-component dark matter model, the expression for the direct detection cross-section can be expressed as,

	\begin{equation}
	\sigma=\left(\dfrac{\Omega_N}{\Omega_{\chi}}\right) \sigma_N + \left(\dfrac{\Omega_{L_S}}{\Omega_{\chi}}\right) \sigma_{L_S}
	\end{equation}
	Individual spin-independent and dependent cross-sections of $N$ and $L_S$, $\sigma_N$ and $\sigma_{L_S}$, have already been discussed in section \ref{subsec31}. So with the knowledge of relative abundances of $N$ and $L_S$ one can easily calculate the direct detection cross-section. As for example, in Fig. \ref{new_eps}, we have presented the spin-dependent scattering cross-section of $N$ (as the contribution of $L_S$ to $\sigma$ will be negligible compared to $N$ even after having a larger relative abundance) as a function of $\epsilon'/\Lambda'^2$ for a specific value of $y_{BB}$ allowed by PLANCK limit. We also compare this cross-section with experimental upper limit from LUX. So in this two-component scenario a new limit on $\epsilon'/\Lambda'^2$ ($5.45-11.96\times 10^{-6}$ GeV$^{-2}$ for $y_{BB}=0.2211$) has been obtained from this analysis. The limits have a little sensitivity on the choice of values of $y_{BB}$ as this parameter controls the co-annihilation of $N$ with $L_S$. It is evident from the plot that the allowed range of $\epsilon'/\Lambda'^2$ is wider than what we obtainned from Fig. \ref{eps}. This can be explained by the reduced relative abundance of $N$ in case of two-component DM.
	
	\section{Conclusion}
	\label{sec5}
	
	To summarise, we have investigated phenomenological implications of a LR symmetric model based on $E_6$ inspired gauge group $SU(3)_C \otimes SU(2)_L \otimes U(1)_L \otimes SU(2)_R \otimes U(1)_R$ from the perspective of Dark Matter. There are two charge neutral fermions among the {\bf 27}-plet of fermions of $E_6$ that we have considered in our analysis. We have investigated in some details the prospect of these two heavy fermions to be a candidate for relic particle. One of these, $N$, is charged under both $SU(2)$s and $U(1)$s transformations. The other one $L_S$ carries $U(1)$ charges however singlet under both the $SU(2)$s. 		
	Among $N$ and $L_S$, $N$ has a larger annihilation rate along with large direct detection cross-section at an experiment like XENON. To bring the direct detection cross-section of $N$ below the experimental limits, we consider a dimension-6 4-fermion effective interaction involving a pair of $N$ and a pair of SM fermions. By appropriately adjusting the coefficient of this operator one can easily make the direct-detection rate of $N$ consistent with the experimental data, however, this will also change the annihilation rate of $N$ in such a way that relic density constraint could not be satisfied.
	
 	We then turn our attention to $L_S$. $L_S$ being an $SU(2)$ singlet, its rate of interaction is smaller than that of $N$ and thus cannot be detected by the present direct detection experiments.
	
	We analyse our case with two different choices of the term $y_{BB}$. $N$ and $L_S$ cannot be treated as an individual relic particle if $y_{BB}=0$. In this case both of $N$ and $L_S$ are stable or non-decaying irrespective of their mass hierarchy and both of them will contribute to the relic density of the Universe. But this two-component DM scenario is not at all promising because of the overabundance of the relic particles (mainly due to $L_S$).
		
	 On the other hand with a non-zero $y_{BB}$, two different scenarios may arise. The first, $N$ and $L_S$ are chosen to be non-degenerate in mass, the heavier between them will decay to the lighter one giving rise to a single component dark matter scenario where the lighter particle will be the relic candidate. Again we cannot arrive at a promising scenario with this choice. If $L_S$ is the relic particle, the scenario yields more than necessary amount of relic and if $N$ is the lightest, it would produce too little amount of relic. 
		
	So the second and only choice we are left with is to set them degenerate or nearly degenerate in mass. This degeneracy will prohibit them from decaying to the other and thus giving rise to a two-component DM scenario. A non-zero $y_{BB}$ will open up the co-annihilation channels between $N$/$E$ and $L_S$ which will reduce the relic density of $L_S$. For a small range of $y_{BB}$ ($0.2 < y_{BB} < 0.236$) would produce about right amount of relic density for a specific range of DM mass. Beyond this range of $y_{BB}$, we shall have over-abundant and under-abundant scenarios respectively.

	However, the relative mass hierarchy among $N$ or $L_S$ and other exotic particles like $q_S$ and $E$ also plays a crucial role in determining the relic density. Making, $L_S$ heavier than $q_S$ (and $E$), would force $L_S$ to annihilate more to $q_S$ and $E$ pairs and will yield a relic density which is in agreement with the experiments. Both $q_S$ and $E$ probably result into SM particles via gauge interactions. However due to the expansion of the Universe, the rate of such annihilations of $q_S$ and $N$ will be reduced and probably result into some remnant $q_S$ or $E$ at present epoch along with $N$ and $L_S$. To avoid such a circumstance, we keep the masses of $q_S$ heavier than $L_S$ and $N$. $E$ having gauge interactions, will probably decay to $N$ along with a SM particle (possibly a $W$) and would contribute to the relic of $N$.
	
	We now turn our attention to the direct detection cross-sections involving $N$ and $L_S$. The effective direct detection cross-section is a combination of the direct detection rates of the individual species weighted by their relative abundances (relic density fractions). Although $L_S$ has a higher relative abundance, its contribution will be negligible compared to its partner $N$ due to its negligibly small direct detection cross-section. 
	It is worth to mention that the direct detection cross-section of $N$ is sensitive to $\epsilon'/\Lambda'^2$, the coefficient of dimension-6 operator, leading to its annihilation to a pair of SM fermions. For a given mass of $N$, we found a range of values of $\epsilon'/\Lambda'^2$ allowed by the direct detection experiment like XENON. 
	
	We have observed that in all possible scenarios that we have considered, rates of indirect detection (DM annihilation to pair of SM particles) of $N$ and $L_S$ is far below than the experimental data from Fermi-LAT. Consequently, one cannot put a meaningful constraint on the model parameters from such an experiment. However, inclusion of Sommerfeld enhancement effect could drastically enhance the annihilation rate and it may be possible to satisfy the experimental data from sattelite experiments. We have not considered such an effect in our analysis.
	
	Before we finally conclude it is interesting to note that the associated production cross-section of a pair of $N$ and $L_S$ with a jet is 0.023 (0.6) fb at 14 (27) TeV LHC run respectively.\\\\

	{\bf Acknowledgement :} S. B. acknowledges financial support from DST, Ministry of Science and Technology, Government of India in the form of an INSPIRE-Senior Research Fellowship. Both of us are grateful to Prof. Amitava Raychaudhuri for an insightful discussion on Majorana fermions.


\begin{thebibliography}{widestlabel}
		
		\bibitem{Planck} PLANCK collaboration, N. Aghanim \emph{et al.}, \emph{Planck 2018 results VI. Cosmological parameters}, \href{ 	https://doi.org/10.1051/0004-6361/201833910}{Astronomy \& Astrophysics 641, A6 (2020)}, arXiv: [\href{https://arxiv.org/pdf/1807.06209.pdf}{1807.06209}].
		
		\bibitem{WMAP} C. L. Bennett \emph{et. al.}, \emph{Nine-Year Wilkinson Microwave Anisotropy Probe (WMAP) Observations: Final Maps and Results}, \href{https://doi.org/10.1088/0067-0049/208/2/20}{Astrophys.J.208 (Oct., 2013) 20}, arXiv: [\href{https://arxiv.org/pdf/1212.5225.pdf}{1212.5225}].
		
		\bibitem{AMS} AMS collaboration, S.-J. Lin, X.-J. Bi, P.-F. Yin and Z.-H. Yu, \emph{Implications for dark matter annihilation from the AMS-02 $\bar{p}/p$ ratio}, arXiv: [\href{https://arxiv.org/pdf/1504.07230.pdf}{1504.07230}].
		
		\bibitem{Pamela} PAMELA collaboration, O. Adriani \emph{et. al.}, \emph{Measurement of the flux of primary cosmic ray antiprotons with energies of 60-MeV to 350-GeV in the PAMELA experiment}, \href{https://doi.org/10.1134/S002136401222002X}{ETP Lett. 96 (2013) 621–627}.
		
		\bibitem{Fermi1} Fermi-LAT collaboration, M. Ackermann \emph{et. al.}, \emph{Searching for Dark Matter Annihilation from Milky Way Dwarf Spheroidal Galaxies with Six Years of Fermi Large Area Telescope Data}, \href{https://doi.org/10.1103/PhysRevLett.115.231301}{Phys. Rev. Lett. 115, 231301 (2015)}, arXiv: [\href{https://arxiv.org/pdf/1503.02641.pdf}{1503.02641}].
		
		\bibitem{Fermi2} Fermi-LAT and MAGIC collaboration, M. L. Ahnen \emph{et. al.}, \emph{Limits to dark matter annihilation cross-section from a combined analysis of MAGIC and Fermi-LAT observations of dwarf satellite galaxies}, \href{https://doi.org/10.1088/1475-7516/2016/02/039}{JCAP 02 (2016) 039}, arXiv: [\href{https://arxiv.org/pdf/1601.06590.pdf}{1601.06590}].
		
		\bibitem{higgs-discovery} ATLAS collaboration, G. Aad \emph{et al.}, \emph{Observation of a new particle in the search for the Standard Model Higgs boson with the ATLAS detector at the LHC}, \href{https://doi.org/10.1016/j.physletb.2012.08.020}{Phys. Lett. B716 (2012) 1-29}, arXiv: [\href{https://arxiv.org/pdf/1207.7214.pdf}{1207.7214}];
		\hspace{5mm} 
		CMS collaboration, S. Chatrchyan \emph{et al.}, \emph{Observation of a New Boson at a Mass of 125 GeV with the CMS Experiment at the LHC}, \href{https://doi.org/ 	10.1016/j.physletb.2012.08.021}{Phys. Lett. B 716 (2012) 30}, arXiv: [\href{https://arxiv.org/pdf/1207.7235.pdf}{1207.7235}].
		
		\bibitem{SUSY-DM} G. Jungman, M. Kamionkowski and K. Griest, \emph{Supersymmetric Dark Matter}, \href{https://doi.org/10.1016/0370-1573(95)00058-5}{Physics Reports 267 (1996) 195-373}, arXiv: [\href{https://arxiv.org/pdf/hep-ph/9506380.pdf}{hep-ph/9506380}].
		
		\bibitem{susy} C. Mu\~{n}oz, \emph{Models of Supersymmetry for Dark Matter}, \href{https://doi.org/10.1051/epjconf/201713601002}{EPJC Conf, 2017}, arXiv: [\href{https://arxiv.org/pdf/1701.05259.pdf}{1701.05259}].
		
		\bibitem{extrad} D. Hooper and S. Profumo, \emph{Dark Matter and Collider Phenomenology of Universal Extra Dimensions}, \href{https://doi.org/10.1016/j.physrep.2007.09.003}{Phys.Rept.453:29-115,2007}, arXiv: [\href{https://arxiv.org/pdf/hep-ph/0701197.pdf}{hep-ph/0701197}].
		
		\bibitem{LHC limits} ATLAS and CMS Collaborations, C. Autermann, \emph{Search for supersymmetry at the LHC}, \href{https://doi.org/10.1051/epjconf/201716401028}{EPJ Web Conf. 164 (2017) 01028}; 
		\hspace{5mm}
		A. Canepa, \emph{Searches for supersymmetry at the Large Hadron Collider}, \href{https://doi.org/10.1016/j.revip.2019.100033}{Reviews in Physics 4 (2019) 100033};
		\hspace{5mm}
		CMS Collaboration, \emph{Search for resonant and nonresonant new phenomena in high-mass dilepton final states at $\sqrt{s} = 13$ TeV}, \href{https://doi.org/10.1007/JHEP07(2021)208}{JHEP07 (2021) 208}, arXiv: [\href{https://arxiv.org/pdf/2103.02708.pdf}{2103.02708}];
		\hspace{5mm}
		CMS Collaboration, \emph{Search for Large Extra Dimensions in the Diphoton Final State at the Large Hadron Collider}, \href{https://doi.org/10.1007/JHEP05(2011)085}{JHEP05 (2011) 085}, arXiv: [\href{https://arxiv.org/pdf/1103.4279.pdf}{1103.4279}]; 
		\hspace{5mm}
		CMS Collaboration, \emph{Search for new physics with dijet angular distributions inproton-proton collisions at $\sqrt{s}=13$ TeV}, \href{https://doi.org/ 	10.1007/JHEP07(2017)013}{JHEP 07 (2017) 013}, arXiv: [\href{https://arxiv.org/pdf/1703.09986.pdf}{1703.09986}].
		
		\bibitem{little-higgs} A. Freitas, P. Schwaller and D. Wyler, \emph{A Little Higgs Model with Exact Dark Matter Parity}, \href{https://doi.org/10.1007/JHEP02(2011)032}{JHEP 0912:027,2009}, arXiv: [\href{https://arxiv.org/pdf/0906.1816.pdf}{0906.1816}].
		
		\bibitem{LRDM} T. Bandyopadhyay and A. Raychaudhuri, \emph{Left-right model with TeV fermionic dark matter and unification}, \href{https://doi.org/10.1016/j.physletb.2017.05.042}{Physics Letters B 771 (2017) 206–212}, arXiv: [\href{https://arxiv.org/pdf/1703.08125.pdf}{1703.08125}];
		\hspace{5mm}
		D. Borah and A. Dasgupta, \emph{Left Right Symmetric Models with a Mixture of keV-TeV Dark Matter}, \href{https://doi.org/10.1088/1361-6471/ab2570}{J. Phys. G: Nucl. Part. Phys. 46 105004}, arXiv: [\href{https://arxiv.org/pdf/1710.06170.pdf}{1710.06170}];
		\hspace{5mm}
		S. Dey, P. Ghosh and S. K. Rai, \emph{Confronting dark fermion with a doubly charged Higgs in the left-right symmetric model}, \href{https://doi.org/10.1140/epjc/s10052-022-10778-z}{Eur. Phys. J. C 82, 876 (2022)}, arXiv: [\href{https://arxiv.org/pdf/2202.11638.pdf}{2202.11638}] and references therein.
		
		\bibitem{extended scalar} R. Campbell, S. Godfrey, H. E. Logan and A. Poulin, \emph{Real singlet scalar dark matter extension of the Georgi-Machacek model}, \href{https://doi.org/10.1103/PhysRevD.95.016005}{Phys. Rev. D 95, 016005 (2017)}, arXiv: [\href{https://arxiv.org/pdf/1610.08097.pdf}{1610.08097}]; \hspace{5mm}
		S. Bhattacharya, P. Ghosh, A. K. Saha, A. Sil, \emph{Two component dark matter with inert Higgs doublet: neutrino mass, high scale validity and collider searches}, \href{https://doi.org/10.1007/JHEP03(2020)090}{J. High Energ. Phys. 2020}, arXiv: [\href{https://arxiv.org/pdf/1905.12583.pdf}{1905.12583}].
		
		\bibitem{extrau1} D. Nanda and D. Borah, \emph{Common origin of neutrino mass and dark matter from anomaly cancellation requirements of a $U(1)_{B-L}$ model}, \href{https://doi.org/10.1103/PhysRevD.96.115014}{Phys. Rev. D 96, 115014}, arXiv: [\href{https://arxiv.org/pdf/1709.08417.pdf}{1709.08417}].
		
		\bibitem{E6} Q. Shafi, \emph{$E_6$ as a unifying gauge symmetry}, \href{https://doi.org/10.1016/0370-2693(78)90248-4}{Physics Letters B, 79(3), 301-303}; 
		\hspace{5mm}
		F. G\"{u}rsey, P. Ramond and P. Sikivie, \emph{A universal gauge theory model based on $E_6$}, \href{https://doi.org/10.1016/0370-2693(76)90417-2}{Physics Letters B, 60(2), 177-180}.
		
		\bibitem{32121} S. Bhattacharyya and A. Datta, \emph{Phenomenology of an $E_6$ inspired extension of Standard Model: Higgs sector}, \href{https://doi.org/10.1103/PhysRevD.105.075021}{Phys. Rev. D 105, 075021}, arXiv: [\href{https://arxiv.org/pdf/2109.08524.pdf}{2109.08524}].
		
		\bibitem{fermion-DM} B. Barman, S. Bhattacharya, P. Ghosh, S. Kadam and N. Sahu, \emph{Fermion dark matter with scalar triplet at direct and collider searches}, \href{https://doi.org/10.1103/PhysRevD.100.015027}{Phys. Rev. D 100, 015027}, arXiv: [\href{https://arxiv.org/pdf/1902.01217.pdf}{1902.01217}]; \hspace{5mm}
		C. Bonilla, L. M. G. de la Vega, J. M. Lamprea, R. A. Lineros and E. Peinado, \emph{Fermion Dark Matter and Radiative Neutrino Masses from Spontaneous Lepton Number Breaking}, \href{https://doi.org/10.1088/1367-2630/ab7254}{New Journal of Physics, Vol. 22, March 2020}, arXiv: [\href{https://arxiv.org/pdf/1908.04276.pdf}{1908.04276}]; \hspace{5mm}
		S. Choubey, S. Khan, M. Mitra and S. Mondal, \emph{Singlet-triplet fermionic dark matter and LHC phenomenology}, \href{https://doi.org/10.1140/epjc/s10052-018-5785-1}{Eur. Phys. J. C 78, 302 (2018)}, arXiv: [\href{https://arxiv.org/pdf/1711.08888.pdf}{1711.08888}]; \hspace{5mm}
		A. Carmona, J. C. Ruiz and M. Neubert, \emph{A warped scalar portal to fermionic dark matter}, \href{https://doi.org/10.1140/epjc/s10052-021-08851-0}{ Eur. Phys. J. C 81, 58 (2021)}, arXiv: [\href{https://arxiv.org/pdf/2011.09492.pdf}{2011.09492}]; \hspace{5mm}
		E. Bernreuther, S. Heeba and F. Kahlhoefer, \emph{Resonant Sub-GeV Dirac Dark Matter}, \href{https://doi.org/10.1088/1475-7516/2021/03/040}{JCAP 2021 (2021) no. 03, 040}, arXiv: [\href{https://arxiv.org/pdf/2010.14522.pdf}{2010.14522}]; \hspace{5mm}
		B. Chauhan, \emph{Sub-MeV Self Interacting Dark Matter}, \href{https://doi.org/10.1103/PhysRevD.97.123017}{Phys. Rev. D 97, 123017 (2018)}, arXiv: [\href{https://arxiv.org/pdf/1711.02970.pdf}{1711.02970}]. 
		
		
		\bibitem{axion} C. D. R. Carvajal, R. Longas, O. Rodr\'{\i}guez and \'{O}. Zapata, \emph{Singlet fermion dark matter and Dirac neutrinos from Peccei-Quinn symmetry}, \href{https://doi.org/10.1103/PhysRevD.105.015003}{Phys. Rev. D. 105, 015003}, arXiv: [\href{https://arxiv.org/pdf/2110.15167.pdf}{2110.15167}].
		
		\bibitem{boltzman} A. Ahmed, M. Duch, B. Grzadkowski and M. Iglicki, \emph{Multi-component dark matter: the vector and fermion case}, \href{https://doi.org/10.1140/epjc/s10052-018-6371-2}{Eur. Phys. J. C (2018) 78:905}, arXiv: [\href{https://arxiv.org/pdf/1710.01853.pdf}{1710.01853}]; \hspace{5mm}
		A. Betancur, G. Palacio and A. Rivera, \emph{Inert doublet as multicomponent dark matter}, \href{https://doi.org/10.1016/j.nuclphysb.2020.115276}{Nucl. Phys. B, 2020, 115276}, arXiv: [\href{https://arxiv.org/pdf/2002.02036.pdf}{2002.02036}]; \hspace{5mm}
		J. Edsjo and P. Gondolo, \emph{Neutralino Relic Density including Coannihilations}, \href{https://doi.org/10.1103/PhysRevD.56.1879}{Phys. Rev. D 56:1879-1894,1997}, arXiv: [\href{https://arxiv.org/pdf/hep-ph/9704361.pdf}{hep-ph/9704361}].
		
		\bibitem{Triparno-Rinku} T. Bandyopadhyay and R. Maji, \emph{The $E_6$ route to multicomponent dark matter}, arXiv: [\href{https://arxiv.org/pdf/1911.13298.pdf}{1911.13298}].
		
		\bibitem{see-saw} R. N. Mahapatra ans P. B. Pal, \emph{Massive Neutrinos in Physics and Astrophysics}, World Scientific Publishing Co. Pte. Ltd., ISBN: 981-238-070-1.
		
		\bibitem{W-mixing} P.A. Zyla \emph{et al.}, \emph{Review of Particle Physics} (Particle Data Group), \href{https://doi.org/10.1093/ptep/ptaa104}{Prog. Theor. Exp. Phys. 2020, 083C01 (2020)}. 
		
		\bibitem{Higgs-invisible} ATLAS collaboration, M. Aaboud \emph{et. al.}, \emph{Combination of searches for invisible Higgs boson decays with the ATLAS experiment}, \href{https://doi.org/10.1103/PhysRevLett.122.231801}{Phys. Rev. Lett. 122, 231801 (2019)}, arXiv: [\href{https://arxiv.org/pdf/1904.05105.pdf}{1904.05105}];
		\hspace{5mm}
		CMS collaboration, A. M. Sirunyan \emph{et. al.}, \emph{Search for invisible decays of a Higgs boson produced through vector boson fusion in proton-proton collisions at $\sqrt{s}=13$ TeV}, \href{https://doi.org/10.1016/j.physletb.2019.04.025}{Phys. Lett. B 793 (2019) 520}, arXiv: [\href{https://arxiv.org/pdf/1809.05937.pdf}{1809.05937}].
		
		\bibitem{scalar-DM1} R. Barbieri, L. J. Hall and V. S. Rychkov, \emph{Improved Naturalness with a Heavy Higgs: An Alternative Road to LHC Physics}, \href{https://doi.org/10.1103/PhysRevD.74.015007}{Phys. Rev. D74:015007 (2006)}, arXiv: [\href{https://arxiv.org/pdf/hep-ph/0603188.pdf}{hep-ph/0603188}].
		
		\bibitem{direct-detection} XENON collaboration, E. Aprile \emph{et al.}, \emph{Dark Matter Search Results from a One Tonne $\times$ Year Exposure of XENON1T}, \href{https://doi.org/10.1103/PhysRevLett.121.111302}{Phys. Rev. Lett. 121, 111302 (2018)}, arXiv: [\href{https://arxiv.org/pdf/1805.12562.pdf}{1805.12562}]; \hspace{5mm} XENON collaboration, E. Aprile \emph{et al.}, \emph{Constraining the Spin-Dependent WIMP-Nucleon Cross Sections with XENON1T}, \href{https://doi.org/10.1103/PhysRevLett.122.141301}{Phys. Rev. Lett. 122, 141301 (2019)}, arXiv: [\href{https://arxiv.org/pdf/1902.03234.pdf}{1902.03234}].
		
		\bibitem{lux} LUX Collaboration, D. S. Akerib \emph{et al.}, \emph{Limits on Spin-Dependent WIMP-Nucleon Cross Section Obtained from the Complete LUX Exposure}, \href{https://doi.org/10.1103/PhysRevLett.118.251302}{PRL 118, 251302 (2017)}, arXiv: [\href{https://arxiv.org/pdf/1705.03380.pdf}{1705.03380}].
		
		\bibitem{pico} PICO Collaboration, C. Amole \emph{et al.}, \emph{Dark Matter Search Results from the PICO-60 $C_3 F_8$ Bubble Chamber}, \href{https://doi.org/10.1103/PhysRevLett.118.251301}{Phys. Rev. Lett. 118, 251301 (2017)}, arXiv: [\href{https://arxiv.org/pdf/1707.07666.pdf}{1707.07666}].
		
		\bibitem{panda} PANDA-II Collaboration, J. Xia \emph{et al.}, \emph{PandaX-II Constraints on Spin-Dependent WIMP-Nucleon Effective Interactions}, \href{https://doi.org/10.1016/j.physletb.2019.02.043}{Physics Letters B 792C (2019) 193-198}, arXiv: [\href{https://arxiv.org/pdf/1807.01936.pdf}{1807.01936}].
		
		\bibitem{DDref} A. Berlin, D. Hooper and S. D. McDermott, \emph{Simplified Dark Matter Models for the Galactic Center Gamma-Ray Excess}, \href{https://doi.org/10.1103/PhysRevD.89.115022}{Phys. Rev. D 89, 115022 (2014)}, arXiv: [\href{https://arxiv.org/pdf/1404.0022.pdf}{1404.0022}].
		\bibitem{slansky} R. Slansky, \emph{Group Theory for Unified Model building}, \href{https://doi.org/10.1016/0370-1573(81)90092-2}{Phys. Rep. 79, 1(1981).}
		
		\bibitem{feynrules} A. Alloul, N. D. Christensen, C. Degrande, C. Duhr and B. Fuks, \emph{FeynRules 2.0 - A complete toolbox for tree-level phenomenology}, \href{https://doi.org/10.1016/j.cpc.2014.04.012}{Comput.Phys.Commun. 185 (2014) 2250-2300}, arXiv: [\href{https://arxiv.org/pdf/1310.1921.pdf}{1310.1921}].
		
		\bibitem{micromegas} G. B\'{e}langer, A. Mjallal and A. Pukhov, \emph{Recasting direct detection limits within micrOMEGAs and implication for non-standard Dark Matter scenarios}, \href{https://doi.org/10.1140/epjc/s10052-021-09012-z}{Eur. Phys. J. C81no. 3, (2021) 239}, arXiv: [\href{https://arxiv.org/pdf/2003.08621.pdf}{2003.08621}].
		
		\bibitem{srednicki} M. Srednicki, R. Watkins and K. A. Olive, \emph{Calculations of Relic Densities in the Early Universe}, \href{https://doi.org/10.1016/0550-3213(88)90099-5}{Nuclear Physics B310 (1988) 693-713}.
		
		\bibitem{subhaditya} S. Bhattacharya, P. Ghosh and P. Poulose, \emph{Multipartite Interacting Scalar Dark Matter in the light of updated LUX data}, \href{https://doi.org/10.1088/1475-7516/2017/04/043}{JCAP 04 (2017) 043}, arXiv: [\href{https://arxiv.org/pdf/1607.08461.pdf}{1607.08461}].
		
		\bibitem{non-thermal} P. S. Dev, A Majumdar and S. Qutub, \emph{Constraining non-thermal and thermal properties of Dark Matter}, \href{https://doi.org/10.3389/fphy.2014.00026}{Front. Phys., 09 May 2014}, arXiv: [\href{https://arxiv.org/pdf/1311.5297.pdf}{1311.5297}].
		
		\bibitem{sommerfeld} N. Arkani-Hamed, D. P. Finkbeiner, T. R. Slatyer and N. Weiner, \emph{A Theory of Dark Matter}, \href{https://doi.org/10.1103/PhysRevD.79.015014
		}{Phys.Rev.D79:015014,2009}, arXiv: [\href{https://arxiv.org/pdf/0810.0713.pdf}{0810.0713}].
	\end{thebibliography}
\end{document}